\documentclass[a4paper,11pt]{article}
\pdfoutput=1 % if your are submitting a pdflatex (i.e. if you have
\usepackage{jheppub} % for details on the use of the package, please
\usepackage[T1]{fontenc} % if needed
\usepackage{amsmath}
\usepackage{amssymb}
\usepackage{amsfonts}
\usepackage{graphicx}
\usepackage{enumitem}
\usepackage{bm}
\usepackage{rotating}
\usepackage{hyperref}
\usepackage{pbox}
\usepackage[dvipsnames]{xcolor}
\usepackage{array}
\usepackage{physics}
\usepackage{dsfont}
\usepackage{mathtools}
\usepackage{leftidx}
\usepackage{lmodern}
\usepackage[normalem]{ulem}
\usepackage{braket}
\usepackage{tensor}
\usepackage{mathrsfs}
\usepackage{float}
\usepackage[title]{appendix}

\usepackage{tikz}
\usetikzlibrary{arrows.meta,decorations.pathmorphing,math}

\newcommand{\sx}{\mathsf{x}}
\newcommand{\sy}{\mathsf{y}}

\newcommand{\rr}[1]{\left(#1\right)}
\newcommand{\openone}{\mathds{1}}

\newcommand{\tin}{\mathrm{in}}
\newcommand{\tout}{\mathrm{out}}

\numberwithin{equation}{section}

\title{Quantum imprints of gravitational shockwaves}
\date{May 20, 2021}

\author[a,b]{Finnian Gray,}
\author[a,b]{David Kubiz\v n\'ak,}
\author[a,b]{Taillte May,}
\author[a,b]{\\Sydney Timmerman,}
\author[b,c]{Erickson Tjoa}

\affiliation[a]{Perimeter Institute for Theoretical Physics, \\ Waterloo, Ontario N2L 2Y5, Canada}
\affiliation[b]{Department of Physics and Astronomy, University of Waterloo,\\ Waterloo, Ontario, N2L 3G1, Canada}
\affiliation[c]{Institute for Quantum Computing, University of Waterloo,\\ Waterloo, Ontario, N2L 3G1, Canada}

\emailAdd{fgray@perimeterinstitute.ca}
\emailAdd{e2tjoa@uwaterloo.ca}
\emailAdd{tmay@perimeterinstitute.ca}
\emailAdd{stimmerman@perimeterinstitute.ca}
\emailAdd{dkubiznak@perimeterinstitute.ca}

\abstract{
Gravitational shockwaves are simple exact solutions of Einstein equations representing the fields of ultrarelativistic sources and idealized gravitational waves (shocks). Historically, much work has focused on shockwaves in the context of possible black hole formation in high energy particle collisions, yet they remain at the forefront of research even today. Representing hard modes in the bulk, shocks give rise to the  gravitational memory effect at the classical level and implant supertranslation (BMS) hair onto a classical spacetime at the quantum level. The aim of this paper is to
further our understanding of the `information content' of such supertranslations. Namely, we show that, contrary to the several claims in the literature, a gravitational shockwave {\em does leave} a quantum imprint on the {\em vacuum state} of a test quantum field and that this imprint is accessible to local observers carrying  Unruh--DeWitt (UDW) 
detectors in this spacetime.
}

\begin{document} 
\maketitle
\flushbottom

\section{Introduction}

Quantum field theory (QFT) in curved spacetime is an infrared (IR) low-energy effective field theory designed to probe certain features of the currently-unknown quantum theory of gravity. It is constructed by lifting standard QFT to curved spacetime, via changing the background metric, and is expected to be valid in the regime where  quantum fluctuations are small enough, so that the semi-classical Einstein field equations sourced by quantum matter are valid. Even in the test field approximation where there is no backreaction to the geometry, QFT in curved spacetime has helped to uncover a number of physical phenomena, such as Unruh and Hawking radiation, and provided firm ground for black hole thermodynamics, while giving rise to one of the greatest puzzles in theoretical physics---the black hole information paradox.

More recent developments have uncovered universal and surprising relationships between the classical gravitational displacement memory effect \cite{Zeldovich:1974gvh,Favata2010memory,Tolish2014memory,Christodoulou1991memory,Winicour2014memory}, leading soft graviton theorems \cite{Weinberg1965graviton,Saha2020softgraviton,Laddha2020softgraviton}, and supertranslation asymptotic symmetries~\cite{sachs1962gravitational,bondi1962gravitational,Strominger2014scattering,Ashtekar1981asymptotic,he2014newQEDsymm,campiglia2015asymptotic,Strominger2014yangmills,Compere2019towermemory,zhang2017memory}: the so-called {\em IR triangle}. One implication of the IR triangle is that the vacuum state of a quantum field in an asymptotically flat background is invariant under the infinite-dimensional Bondi--Metzner--Sachs (BMS) group, which contains the Poincar\'e group as a finite-dimensional subgroup. The BMS group was investigated in the 1960s in an attempt to better understand gravitational radiation \cite{sachs1962gravitational,bondi1962gravitational}, but it was not until recently that the connection to QFT has become more apparent. 

To demonstrate this connection one needs to have a physical mechanism for implanting a {\em supertranslation hair} onto a classical spacetime. This can be achieved with a gravitational {\em shockwave} \cite{Hawking:2016sgy}, which is a localized (distributional) gravitational wave that coincides with the background (without the wave) everywhere except at the wavefront (shock) located at a single value of the null coordinate. The shockwave is sourced by a singular stress energy tensor along the null direction and can be described by a series of exact solutions of the Einstein equations going back to the Aichelburg--Sexl metric~\cite{Aichelburg:1970dh}, and, more generally, the Dray and 't Hooft solution~\cite{Dray1985shockwave,tHooft:1987vrq}. Such metrics have played an important role in studying gravitational scattering and the conditions for black hole formation in high energy particle collisions~ \cite{tHooft:1987vrq,deVega:1988wp,Amati:1992zb,Kabat:1992tb,Eardley:2002re}, and as discussed in~\cite{Kolekar2017memoryaccel,Shore2018memory,Donnay2018memory}, they give rise to the classical gravitational memory effect. Moreover, shockwaves are at the forefront of research even today, e.g.~\cite{Liu:2021kay}, and also have implications for the black hole information paradox~\cite{Strominger:2017aeh,Gaddam:2020mwe}.  In this context the following two natural questions arise: are there any quantum imprints of gravitational shockwaves (or supertranslation  hairs) on test quantum scalar fields living on the background geometry? And, if so, are they readily accessible by localized observers?

Similar questions have been investigated by a number of recent studies \cite{Kolekar2018quantummemory,Compere:2019rof,Majhi2020shock,Ferreira:2020whz}.
Our work is mainly motivated by \cite{Compere:2019rof,Majhi2020shock} where the authors studied the impact of a `matter induced' supertranslation %at null infinity
on the Bogoliubov coefficients between the two asymptotic states.  Therein, the Bogoliubov coefficients were shown to have physical manifestation only for non-vacuum states of the field; in particular, it was shown that the vacuum expectation values of the (global) number operators associated to both Minkowski and Rindler observers are not altered by the shockwave.
However, since there is no localized number operator that agrees with the global number operator for the vacuum state, the Bogoliubov coefficients are arguably not a true local observable in this context, a result deeply connected to the Reeh--Schlieder theorem, see \cite{Dowker:2011ga,Kubicki:2016ngz,Colosi2008localization,Licht1963localization,Witten2018entanglement,Strohmaier2002curvedRS} for further discussions. Therefore, the results obtained in \cite{Kolekar2018quantummemory,Compere:2019rof,Ferreira:2020whz} are global results that no localized observers can easily access. The method employed in \cite{Majhi2020shock} is  more suitable for local calculations, since it relies on a computation of the Wightman function of the quantum field on the shockwave background, which is known to be accessible to local observers,  e.g. \cite{Jose2018measure}.  However, we shall show in this paper that the study in \cite{Majhi2020shock} is incomplete, as the Wightman function was only calculated for spacetime events located on the same side of the shockwave. 

In summary, the studies \cite{Compere:2019rof,Majhi2020shock,Ferreira:2020whz} have reached the same conclusion: for asymptotic vacua both Unruh and Hawking spectra are unchanged by the presence of gravitational shockwaves, and, while gravitational shockwaves can leave imprints on test quantum fields, this will only happen for non-vacuum states. It is the purpose of the present paper to refute this conclusion.

Intimately tied to the notion of local and \emph{measureable} observables in QFT is the Unruh--Dewitt (UDW) detector. The detector models the local measurement of a QFT via its interaction with a two-level non-relativistic quantum system, a qubit~\cite{Unruh1979evaporation, DeWitt1979}. This simple setup 
is known to reproduce essential aspects of light-matter interactions \cite{pozas2016entanglement,Lopp2021deloc}.
The UDW detector has also been shown to be sensitive to 
global properties of the spacetime as well as local curvature effects, including the passage of gravitational waves~\cite{smith2016topology,Smith2020harvestingGW,VerSteeg2009entangling,Tjoa2020vaidya,henderson2018harvestingBH,henderson2019AdSharvesting,cong2019entanglement,Cong2021inertialdrag}. New effects such as the anti-Unruh and anti-Hawking effects, only accessible to local and finite time interaction observers, have also been observed within this framework~\cite{Garay2016anti-unruh,Brenna2016anti-unruh,Henderson2020anti-hawking,Dappiaggi2021anti-hawking}.

In this paper we show that, contrary to the conclusions of \cite{Compere:2019rof,Majhi2020shock,Ferreira:2020whz}, a gravitational shockwave \emph{does} leave a quantum imprint on the vacuum state of a test quantum scalar field,
and, this imprint is accessible to local observers carrying UDW detectors
in the shockwave spacetime. Namely, we show that the Wightman function contains an additional term that depends on the shockwave profile and that this term is nonzero between two spacetime events that are located on different sides of the shockwave null plane. Using the entanglement harvesting protocol from relativistic quantum information \cite{pozas2015harvesting,pozas2016entanglement,Valentini1991nonlocalcorr,reznik2003entanglement}, we show that this additional term has operational consequences: two UDW detectors can harvest \textit{more} entanglement from the quantum vacuum even though locally each detector sees a strictly Minkowski vacuum (the excitation probability is equal to the one in Minkowski space). In fact, we demonstrate that even a single UDW detector can see the shockwave as it passes by. This is in contrast to the recent study \cite{Smith2020harvestingGW} where it was shown that a single UDW detector cannot detect (linearized) gravitational waves. As we shall discuss, this negative result can be traced to the fact that the quantum detection was only investigated at the linear order in the wave amplitude, while the measurable effect appears at the second order.

Our paper is organized as follows. In Sec.~\ref{sec:shockwaves} we introduce the classical geometry of shockwave spacetime. In Sec.~\ref{sec:QFT} we review quantum field theory on the shockwave background, and construct the corresponding Wightman function (to be contrasted with partial results of \cite{Majhi2020shock}). Sec.~\ref{sec: UDW} reviews the Unruh--DeWitt detector formalism  and the entanglement harvesting protocol. In Sec.~\ref{sec:results} we present our results regarding the single detector response to the gravitational shockwave and the associated entanglement harvesting. Sec.~\ref{sec:conclusions} is devoted to the final discussion and conclusions. The technical Appendix~\ref{app:A} contains the derivation of the Wightman function, while Appendix~\ref{app:B} describes geodesics in the shockwave spacetime. In this paper we adopt the mostly-plus signature for the metric and 
use natural units $\hbar=c=1$, while keeping the gravitational constant $G$ explicit.

\section{Gravitational shockwaves}
\label{sec:shockwaves}

In what follows we shall consider the Dray and 't Hooft~\cite{Dray1985shockwave} generalization of the Aichelburg--Sexl~\cite{Aichelburg:1970dh} shockwave spacetime in $D$ spacetime dimensions. For a wave propagating in the $z$-direction, the corresponding Brinkmann form (see e.g.~\cite{zhang2017memory}) of the metric reads
\begin{equation}
    \label{eq:metric}
    ds^2 = -d u d v +f(\vec{x}) \delta(u-u_0) d u^2 + \delta_{ij} d x^{i} d x^{j}\,.
\end{equation}
Here, we have employed `Minkowski'-like coordinates $\sx\equiv (t,z,\vec{x})$: $u=t-z$ and $v=t+z$ are the `standard' null coordinates. $\vec{x}$ or $x^i$, $i\in\{2,\dots,D-2\}$, denote the transverse directions (coordinates on the wavefront). The wavefront is localised at $u=u_0$; on either side of $u_0$ the spacetime is exactly Minkowski. The stress-energy tensor has only one non-zero component:
\begin{equation}\label{eq: Energy}
    T_{uu} = \delta(u - u_0)\rho(\vec{x})\,, 
\end{equation} 
and is zero everywhere but at $u=u_0$. 

The Einstein field equations reduce to
\begin{equation}
    \Delta f(\vec{x}) = -16\pi G \rho(\vec{x}) \,,
\end{equation}
where $\Delta=\delta^{ij}\partial_i\partial_j$ is the flat Laplacian in the transverse direction. The shockwave profile $f(\vec{x})$ completely determines the nature of the wave. The only restrictions on $f(\vec{x})$ are that it is `smooth enough' and that suitable energy conditions are satisfied (see e.g.~\cite{Kontou:2020bta,Curiel:2014zba}).  For concreteness, we shall impose the null energy condition which implies $\rho(\vec{x})\geq0$, i.e., $\Delta f(\vec{x})\leq0$.

The metric \eqref{eq:metric} describes a large variety of physical spacetimes. In the original work of Aichelburg--Sexl~\cite{Aichelburg:1970dh} the metric was obtained by `boosting' the Schwarzschild spacetime to the speed of light while keeping the energy constant, thus obtaining a gravitational field of a massless point particle characterized by the shockwave profile
\begin{equation}\label{eq:AS_profile}
    f(\vec{x})=-4GP\log\left(\frac{|{\vec x}|}{x_0}\right)\,,
\end{equation}
where $P>0$ corresponds to the energy of the point particle and $x_0$ is some arbitrary reference scale.

Another interesting example of a shockwave profile (form factor) is given by
\begin{equation}\label{eq:Gauss_profile}
 f(\vec{x})=-\vec{x}\cdot A\cdot \vec{x}=-\sum_{i} a_i (x^i)^2\,,
\end{equation}
where $A$ is a symmetric constant matrix, where, without loss of generality, we can work in transverse coordinates corresponding to the eigenvectors with eigenvalues $a_{i}$ of $A$. Physically this form factor corresponds to a  domain wall boosted to the speed of light~\cite{Vilenkin:1984hy,Lousto:1990wn,Barrabes:2002hn}, while the energy density of the shockwave is\footnote{Before we proceed further let us pause and make a few comments on dimensional analysis. In our units $G=1/m_{pl}^2$, so that $[G]=-2$. Given its relation, \eqref{eq: Energy}, to the stress-energy tensor the dimensions of $\rho$ satisfy $[\rho]+ [\delta]=4$, and since $[\delta]=1$, we have $[\rho]=3$ and $[f]=-1$. Thus \eqref{eq: stress-energy} means that the matrix $A$ has dimensions $[A]=1=\text{Energy}=1/\text{Length}$.}
\begin{equation}  
    \rho =  \frac{\Tr(A)}{8\pi G}\,.
    \label{eq: stress-energy}
\end{equation}
Obviously,  when $\Tr(A)=0$, the corresponding energy momentum tensor vanishes, and we have a sourceless (arbitrarily polarized) 
gravitational wave.

Another perspective on shockwaves, due to Penrose, views them as a `scissor and paste' of two copies of Minkowski spacetime~\cite{Penrose:1972xrn}. This is easiest to understand by introducing a new coordinates $\hat{v}$ related to the old by a `planar supertranslation'~\cite{Compere:2019rof}
\begin{equation}
    \hat{v}=v-\Theta(u-u_0)f(\vec{x})\,,
\end{equation}
where $\Theta$ is the Heaviside step function.
In these coordinates the metric \eqref{eq:metric} takes the form
\begin{align}
    d  s^2 &=-d  u d \hat{v} -\Theta(u-u_0)\partial_i f(\vec{x})\, d  u d  x^i +d \vec{x}^2\nonumber \\
    &= \eta +\Theta(u-u_0) {\cal L}_\xi \eta\,,
    \label{eq: supertranslated-Minkowski}
\end{align}
where $\eta=-d  u d\hat v+d \vec{x}^2$ is the flat metric, and 
\begin{equation}
\xi=f(\vec{x})\partial_{\hat{v}}    
\end{equation}
is the {\em supertranslation vector field}, which reduces to a time-translation vector field when $f(\vec{x})$ is independent of $\vec{x}$. The form \eqref{eq: supertranslated-Minkowski} makes it clear that the Minkowski spacetimes are diffeomorphic since $\xi$ implements a diffeomorphism. The key difference between the shockwave spacetime and more general supertranslations typically 
considered in the IR triangle program (see e.g. \cite{strominger2018lectures}) is that shockwaves correspond to `hard' (finite energy, or equivalently, matter-induced) processes in the bulk, rather than the `soft' (zero energy) supertranslations \cite{Hawking:2016sgy,Donnay2018memory,Compere:2019rof}.

For completeness, let us mention that, in $D=4$, the supertranslation manifests itself as a change in the gravitational radiation data~\cite{strominger2018lectures}. This supertranslation provides an infinitely degenerate label for the Minkowski space---two Minkowski spacetimes labeled by different supertranslations are physically inequivalent from the perspective of gravitational scattering problem. In our case, two vacuum states defined on the `in' and `out' Minkowski regions can be unitarily equivalent in the Poincar\'e sense but nonetheless physically distinct in the BMS sense (see e.g. \cite{campiglia2015asymptotic,he2014newQEDsymm,Strominger2014yangmills}). 

We close this section by briefly commenting on the geodesics in this geometry. It is well-known that geodesics in shockwave geometry are highly non-trivial due to the distributional nature of the $\delta$-localized shockwave. Much work has been done to address this from both physical and mathematical perspectives for arbitrary shockwave profiles, see e.g.  \cite{ferrari1988beam,balasin1997geodesics,kunzinger1999rigorous}. As we are interested in localized observers who carry particle detectors (to be discussed in Section~\ref{sec: UDW}), we will need to calculate the timelike geodesics on this background. A key feature of such  geodesics is the distributional nature of the trajectories that crucially depend on the impact parameter.
As shown in Appendix~\ref{app:B}, for the shockwave profile we consider in the subsequent sections, there is a natural choice of impact parameter that `eliminates' the need to deal with these complications.

\section{Quantum field theory in shockwave geometry} \label{sec:QFT}

Having described the classical aspects of shockwave geometry we now turn to review the quantization of a massless scalar field in this spacetime. While quantum fields in gravitational wave spacetimes have been studied since the 70's, e.g. \cite{Gibbons:1975jb,Garriga:1990dp}, QFT in the shockwave geometry was first studied by  Klimčík~\cite{Klimcik1988shock} and more recently in~\cite{Compere:2019rof,Majhi2020shock}. In this section, we reproduce the essential features for shockwaves and explicitly calculate the two-point (Wightman) function for an arbitrary profile $f(\vec{x})$.

\subsection{Klein--Gordon equation and its solutions}

Following~\cite{Klimcik1988shock}, we consider the massless Klein--Gordon equation 
\begin{equation}\label{wave}
\Box \Phi=0\,,    
\end{equation}
find its mode functions, and canonically quantize them in null %(or light-cone) 
coordinates. Given that $\partial_v$ is a Killing vector of \eqref{eq:metric} we can make the following separation ansatz for plane wave modes in the $\partial_v$ direction: $\phi_{k_-}=e^{-ik_-v} \psi(u,\vec{x})$, labeling the momentum with the wavevector $k_\mu=(k_t,k_z,\vec{k})$ and defining $k_\pm=\frac{1}{2}(k_t\pm k_z)$. This reduces the wave equation \eqref{wave} to the following Schr\"odinger-like equation:
    \begin{equation}\label{eq:Schr}
        i \partial_u \psi = - \left( \frac{\Delta}{4 k_-} + f(\vec{x}) k_- \delta(u-u_0) \right)  \psi\,.
    \end{equation}

Clearly, on either side of the shockwave (in the `in' ($u<u_0$) and `out' ($u>u_0$) regions) the solutions are simple plane waves that provide a complete basis for quantization. Thus, we have two mode expansions at our disposal $\phi^\tin_{k_-,\vec{k}}$ and $\phi^\tout_{k_-,\vec{k}}$, for which we require that the mode functions reduce to plane waves in the `in' or `out' regions, respectively. 

Let us first consider the `in' modes. Before the shockwave we have
the standard plane wave solution
\begin{equation}
     \phi^\tin_{k_-,\vec{k}}\Big|_{u<u_0}=e^{-ik_-v} \psi_<(u,\vec{x})\,, \quad \psi_{<} (u,\vec{x})=N_{k_-}e^{-i(k_+(u-u_0)-\vec{k}\cdot\vec{x}))}\,,
\end{equation}
where $N_{k_-}=([2\pi]^{D-1}2k_-)^{-1/2}$ and $k_+=\vec{k}^2/(4 k_-)$. 
This remains a solution until $u=u_0^-$. The only effect of the shockwave on the mode functions is to introduce a junction condition at $u=u_0$ (much like the `scissor and paste' idea on the spacetime itself). Right after the shockwave, one finds~\cite{Klimcik1988shock}
\begin{equation}\label{3.4}
    \phi^\tin_{k_-,\vec{k}}\,\Big|_{u=u_0^+}=N_{k_-}e^{-ik_-v}e^{i(\vec{k}\cdot\vec{x}+ k_-f(\vec{x}))}\,,
\end{equation}
which can be derived by regularizing the Dirac $\delta$-function.  
In other words, the shockwave has just `supertranslated' the solution, changing $v\rightarrow v-f(\vec{x})$. Eq.~\eqref{3.4} provides an initial condition for the time evolution via the Schr\"odinger-like equation \eqref{eq:Schr}, {which for $u>u_0^+$ yields} (see \cite{Klimcik1988shock} for details)
\begin{equation}
    \psi_>(u,\vec{x})= N_{k_-} e^{-ik_-v}e^{i\vec{k}\cdot\vec{x}}\int \frac{d\vec{k'} d\vec{x'}}{(2\pi)^{D-2}}\, e^{i(\vec{k}-\vec{k'})(\vec{x'}-\vec{x})} e^{\frac{-i \vec{k'}^2}{4k_-}(u - u_0) +ik_-f(\vec{x'})}\,.
\end{equation}
Putting this together, we can write the full `in' mode as\footnote{In these expressions we are implicitly ignoring the zero measure set of modes localized to travel along the shockwave $u=u_0$.}
\begin{align}\label{eq: u_in}
         \phi^\tin_{k_-,\vec{k}}=N_{k_-}e^{-ik_-v}e^{i\vec{k}\cdot\vec{x}}\int \frac{d\vec{x'}d\vec{k'}}{(2\pi)^{D-2}}e^{i(\vec{k}-\vec{k'})(\vec{x'}-\vec{x})-\frac{i\vec{k'}^2}{4k_-}(u-u_0)+ik_-\Theta(u-u_0)f(x')}\,.
\end{align}

In order to obtain the `out' modes we use the fact the Schr\"odinger equation \eqref{eq:Schr} is symmetric under~\cite{Compere:2019rof}
    \begin{equation}
        u\rightarrow-u,\qquad v\rightarrow-v,\qquad k_-\rightarrow-k_-,\qquad k_+\rightarrow -k_+,\qquad u_0\rightarrow-u_0.
    \end{equation}
Applying this yields for the `out' mode 
\begin{align}\label{eq: u_out}
         \phi^\tout_{k_-,\vec{k}}=N_{k_-}e^{-ik_-v}e^{i\vec{k}\cdot\vec{x}}\int \frac{d\vec{x'}d\vec{k'}}{(2\pi)^{D-2}}e^{i(\vec{k}-\vec{k'})(\vec{x'}-\vec{x})-\frac{i\vec{k'}^2}{4k_-}(u-u_0)-ik_-\Theta(u_0-u)f(x')}\,.
\end{align}

\subsection{Mode decomposition} 

It can be shown~\cite{Compere:2019rof} that
the two sets of mode functions \eqref{eq: u_in} and \eqref{eq: u_out} are orthonormal with respect to the standard Klein--Gordon inner product evaluated on a $t=const.$ Cauchy hypersurface: 
\begin{align}
     \left(\phi_1,\phi_2\right) &=i\int_\Sigma d\Sigma^\mu \left(\phi_1^* \partial_\mu \phi_2-  \phi_2 \partial_\mu \phi_1^*\right) \nonumber\\
     &=i\int_\Sigma dz d{\vec x}\Bigl( \phi_1^*(\partial_u +[1+2\delta(u-u_0)f(\vec{x})]\partial_v)\phi_2 -(1\leftrightarrow2)\Bigr)\,.
\end{align}
That is, for $A\in\{\tin,\tout\}$ we have\footnote{Note that we could normalize with respect to $k_t=\omega$ by redefining $N_{k_-}\mapsto N_{k_t}=\sqrt{(k_-/k_t)}N_{k_-}$. }
\begin{align}
    \left(\phi^{A*}_{k_-,\vec{k}},\, \phi^A_{l_-,\vec{l}}\right)&=0\,,\hspace{0.5cm}
     \left(\phi^{A}_{k_-,\vec{k}},\, \phi^A_{l_-,\vec{l}}\right)=\left(\phi^{A*}_{k_-,\vec{k}},\, \phi^{A*}_{l_-,\vec{l}}\right)=\delta(k_- - l_-)\delta(\vec{k}-\vec{l})\,.
\end{align}
Thus we have two equivalent quantizations based of the global decomposition using `in' and `out' modes
\begin{align}\label{eq: ModeExp}
    \Phi = \int dk_-d\vec{k}\, \left(a_{k_-,\vec{k}}^\tin\, \phi_{k_-,\vec{k}}^\tin + a_{k_-,\vec{k}}^{\tin \dagger}\,\phi_{k_-,\vec{k}}^{\tin *}\right) = \int dk_-d\vec{k}  \left(a_{k_-,\vec{k}}^\tout\,\phi_{k_-,\vec{k}}^\tout + a_{k_-,\vec{k}}^{\tout \dagger}\, \phi_{k_-,\vec{k}}^{\tout *}\right)\,,
\end{align}
where the annihilation operators define the `in' and `out' vacua, $\ket{0_\tin}$ and $\ket{0_\tout}$, respectively:
\begin{align}
    a_{k_-,\vec{k}}^\tin \ket{0_\tin}\equiv0\,,\qquad a_{k_-,\vec{k}}^\tout \ket{0_\tout}\equiv0\,.
\end{align}
These operators of course obey the standard commutation relations
\begin{equation}
    \left[ a_{k_-,\vec{k}}^{A}, a_{l_-,\vec{l}}^{A\,\dagger}\right]=\delta(k_- - l_-)\delta(\vec{k}-\vec{l})\,
\end{equation}
with all others vanishing. 

\subsection{Bogoliubov coefficients}

Now we want to find a relation between $\phi^\tin_{k_-,\vec{k}}$ and $\phi^\tout_{k_-,\vec{k}}$. In fact, since the positive and negative frequencies are not mixed, c.f. \eqref{eq: u_in} and \eqref{eq: u_out}, we have
\begin{equation}
    \phi^\tin_{k_-,\vec{k}}=\int d l_- d\vec{l}\; \alpha_{k_- , l_-} (\vec{k}-\vec{l})\; \phi^\tout_{l_-,\vec{l}}\;\,.
\end{equation}
Equivalently this implies that the creation/annihilation operators in the two vacua are related by
\begin{equation}
    a^\tout_{k_{-},\vec{k}}=\int dl_-\,d\vec{l}\, \,\alpha_{k_-,l_-}(\vec{k}-\vec{l})\, a^\tin_{l_{-},\vec{l}}\;\,.
\end{equation}
Thus, we only have one set of Bogoliubov coefficients $ \alpha_{k_- ,l_-}(\vec{k}-\vec{l})$  which means there is no particle production induced by the shockwave in the sense that,
\begin{equation}
   \bra{0_\tin} n^\tout_{k_-,\vec{k}} \ket{0_\tin}= \bra{0_\tin}a^{\tout\dagger}_{k_-,\vec{k}} \,a^\tout_{k_-,\vec{k}}\ket{0_\tin}=0\,.
\end{equation}
This calculation means that the two `Minkowski' vacua defined on the `in' and `out' regions are unitarily equivalent.

In ~\cite{Compere:2019rof} these statements were generalized to consider the shockwave in Rindler coordinates. Therein, it was found that the shockwave only affected the phase of the Bogoliubov coefficients between the Rindler vacuum and the stationary vaccuum as defined here. Hence the Unruh radiation spectrum is left invariant under these shockwaves---only excited states directly see the effect of the profile. Similar results were also obtained for Hawking radiation, see \cite{Ferreira:2020whz}.

To find the Bogoliubov coefficients explicitly, we note that, since $\partial_v$ is a Killing vector, there can be no change of momentum in the $v$ direction. That is 
\begin{equation}
    \alpha_{k_- ,l_-}(\vec{k}-\vec{l})=\delta(k_- - l_-)\, \widetilde{\alpha}_{k_-}(\vec{k}-\vec{l})\,,
\end{equation}
where $\widetilde{\alpha}$ is known as the mixing factor. Comparing the `in' and `out' modes yields
\begin{align}
      \widetilde{\alpha}_{k_-}(\vec{k}-\vec{l}) \equiv \int \frac{d\vec{x'}}{(2\pi)^{D-2}} e^{ i(\vec{k} -\vec{l} )\vec{x}' +i k_- [\Theta(u-u_0)-\Theta(u_0-u)]f(\vec{x}')}\,.
\end{align}

While these statements certainly hold with regards to the creation and annihilation operators and the number operator $N^A_k=a^{A\,\dagger}_{k_{-}, \vec{k}} a^{A }_{k_{-}, \vec{k}}$, it is not clear what this means for the local observers. To answer this we will calculate the Wightman function, which plays a central role in the Unruh--DeWitt paradigm, as outlined in 
Sec.~\ref{sec: UDW}.

\subsection{Wightman function}

The Wightman function was calculated in \cite{Majhi2020shock} on either side of the shockwave, and was found to reduce to the standard Minkowski one. However, in \cite{Majhi2020shock}, the most interesting case---with the events on opposite sides of the shockwave---was omitted. In this section we will calculate the Wightman function in the full spacetime. Using the mode expansion \eqref{eq: ModeExp} and assuming that we are in the `in' vacuum state, we have
\begin{align}\label{eq: W_Def}
    W(\sx,\sy)=\bra{0_\tin}\Phi(\sx)\Phi(\sy)\ket{0_\tin}=\int dk_-d\vec{k} \,\phi^\tin_{k_{-},\vec{k}}(\sx)\,\phi^{\tin\,*}_{k_{-},\vec{k}}(\sy)\,.
\end{align}

Denoting $\sx=(u,v,\vec{x})$ and $\sy=(U,V,\vec{X})$ and plugging in the expressions for the mode functions~\eqref{eq: u_in}, we obtain
\begin{align}\label{eq:WStart}
    W(\sx,\sy)&= \int \frac{dk_{-}\,d\vec{k}}{2(2\pi)^{D-1}k_{-}}e^{-ik_{-}(\Delta v-i\epsilon)}\int\frac{d\vec{x}'d\vec{X}'}{(2\pi)^{2D-4}}e^{i\vec{k}\cdot (\vec{x}'-\vec{X}')+ik_{-}\left[\Theta_uf(\vec{x}')-\Theta_Uf(\vec{X}')\right]}\notag\\
    &\hspace{0.5cm}\times\frac{(4\pi k_-)^{D-2}}{[(u-u_0-i\epsilon)(U-u_0+i\epsilon)]^{\frac{D-2}{2}}}\exp\left[\frac{ik_-(\vec{x}-\vec{x}')^2}{u-u_0-i\epsilon}-\frac{ik_-(\vec{X}-\vec{X}')^2}{U-u_0+i\epsilon}\right]\,,
\end{align}
where we have used the shorthands $\Delta v\equiv v-V$ and $\Theta_u\equiv \Theta(u-u_0)$, and have performed one Gaussian integral along $\vec{k}'$ for $\phi^\tin_{k_-,\vec{k}}$. We have also introduced an $i\epsilon$ to avoid a branch cut. Relegating the rest of the calculation to Appendix \ref{app:A}, the end result is
\begin{align}
    W(\sx,\sy)&= \frac{\Gamma (D-2)(i)^{2-D}}{2(2\pi)^{D-1}}\frac{(4\pi)^{D-2}}{[(u-u_0-i\epsilon)(U-u_0+i\epsilon)]^{\frac{D-2}{2}}} \notag\\
    &\times \int\frac{d\vec{x}'}{(2\pi)^{D-2}}\left(\left[\Delta v-i\epsilon-f(\vec{x}') \Delta\Theta_u\right]- \frac{(\vec{x}-\vec{x}')^2}{u-u_0-i\epsilon}+\frac{(\vec{X}-\vec{x}')^2}{U-u_0+i\epsilon}\right)^{2-D}\,,
    \label{eq:finalW}
\end{align}
where we have defined $\Delta\Theta_u:=\Theta_u-\Theta_U$. In Appendix \ref{app:A} we also show that this reduces to the standard flat space form if either $f(\vec{x})=0$ or $\Delta\Theta_u=0$. In the former there is no shockwave at all, while the latter means that the events $u$ and $U$ are localized at the same side of the shockwave.

This is as far we can go without picking a particular shockwave profile. For simplicity we shall focus on the planar form in \eqref{eq:Gauss_profile}. In this case the integrals can be performed analytically (see Appendix \ref{app:A}) and we obtain
\begin{align}
     W(\sx,\sy) &=\frac{(-i)^{D-2}\Gamma(D/2-1)}{4\pi^{D/2}} \prod_{i=1}^{D-2}\left(1+a_i\Delta\Theta_u\frac{(u-u_0-i\epsilon)(U-u_0+i\epsilon)}{\Delta u-i\epsilon}\right)^{-1/2}\nonumber\\
    &\hspace{-1.65cm}\times\left((\Delta v-i\epsilon)(\Delta u-i\epsilon) -\Delta \vec{x}\,^2 + \sum_{i=1}^{D-2}\frac{a_i\Delta\Theta_u\left([U-u_0+i\epsilon]x^i-[u-u_0-i\epsilon]X^i\right)^2}{\Delta u-i\epsilon+a_i\Delta\Theta_u(u-u_0-i\epsilon)(U-u_0+i\epsilon) }\right)^{(2-D)/2}\!\!\!\!\!\!\!\!\!\!\!\!\!\!\!\!\!\!.
    \label{eq: Wightman-Gauss}
\end{align}
This expression should be contrasted to \cite{Majhi2020shock}, where the Wightman function was not evaluated \textit{across} the shockwave and thus all the components that depend on $\Delta\Theta_u$ drop out. Consequently, the main claim of \cite{Majhi2020shock} does not directly corroborate the result in \cite{Compere:2019rof} which properly accounts for observer trajectories that cross the shock. One of the reasons why the vacuum Wightman function encodes information about the shock and the number operator calculations do not, is that the latter neglect the spatial vacuum entanglement between the two regions of the spacetime separated by the shock, which is an intrinsic property of the vacuum for any relativistic QFT \cite{summers1985bell,summers1987bell}.

As we shall see, the Wightman function \eqref{eq: Wightman-Gauss} will allow us to 
detect the passing shockwave with a single UDW detector, and show that two localized quantum systems (UDW detectors) interacting with shockwave vacuum can extract more entanglement than from a global flat space, even though each detector is completely localized in the respective Minkowski region.

\section{Unruh--DeWitt detectors and entanglement harvesting}
\label{sec: UDW}
In this section we will briefly outline the basic framework of  the Unruh--DeWitt (UDW) particle detector and the entanglement harvesting protocol from relativistic quantum information, as well as discuss the `experimental setup' for the quantum detection of the gravitational shockwave by a single UDW detector and by entanglement harvesting.

\subsection{Single UDW detector}

A UDW detector is a pointlike  two-level quantum system (a qubit) interacting locally with an underlying quantum field along its trajectory in spacetime~\cite{Unruh1979evaporation,DeWitt1979}. It provides a simplified model of light-matter interaction where the usual atomic dipole coupled to an electromagnetic field is replaced with a monopole interaction with a scalar field. This approximation is good as long as there is no exchange of angular momentum involved~\cite{pozas2016entanglement}. The pointlike model has the advantage of being simple and covariant---the excitation probabilities are invariant under arbitrary diffeomorphisms~\cite{Tales2020GRQO,Bruno2020time-ordering}.

A single UDW detector couples locally with a scalar field $\Phi$ via the interaction Hamiltonian
\begin{align}
    H_{I}(\tau) &= \lambda\chi(\tau) \mu (\tau)\otimes\Phi(\sx[\tau])\,, 
\end{align}
where $\lambda$ is the coupling constant, $\tau$ is the proper time of the detector, $\sx[\tau]$ is its spacetime trajectory, and $\chi(\tau)$ is the switching function governing the duration of interaction. The operator $\mu$ is the monopole moment of the detector given by
\begin{align}
    \mu(\tau) = \ket{e}\!\bra{g}e^{i\Omega \tau}+\ket{g}\!\bra{e}e^{-i\Omega \tau}\,,
\end{align}
where $\ket{g}$ and $\ket{e}$ are the ground and excited states of the detector, separated by an energy gap $\Omega$. We will assume that the detector switching function is a Gaussian function
\begin{align}\label{eq:Gauss_Switch}
    \chi(\tau) &= \exp\left[{-\frac{(\tau-{\tau_{0})^2}}{T^2}}\right]\,,
\end{align}
where the Gaussian width $T$ controls the duration of interaction and $\tau_{0}$ is the time when the interaction is strongest.\footnote{For our purposes we can think of the detector as being effectively turned on within the time interval $[-bT+\tau_0,bT + \tau_0]$, where $b>0$ is sufficiently large. In fact, already for $b=2$, one recovers $99.5\%$ of the Gaussian area. The compact switching can thus be well-approximated by truncating the exponential tails at $\pm bT+\tau_0$ for some finite $b\geq 2$.}

We can perform a measurement with the UDW detector by preparing it in an initial state, time-evolving it together with the field using the interaction Hamiltonian, and computing the reduced density matrix.  We begin by preparing the detector-field system in the product state
\begin{align}
    \rho_0 &= \ket{g}\!\bra{g}\otimes\ket{0}\!\bra{0}\,,
\end{align}
where $\ket{0}$ is the vacuum state of the field in the asymptotic past.\footnote{ 
In our setup, the initial vacuum state $\ket{0}$ is unitarily equivalent to the vacuum state $\ket{0'}$ in the asymptotic future with respect to the Poincar\'e subgroup, as the Bogoliubov coefficients of both vacua have vanishing `beta coefficients', see previous section. However, since the gravitational shockwave implants supertranslation `hair', we can think of this as the soft charges providing `soft' quantum numbers $\alpha$ (which is uncountably degenerate), and thence the two `Minkowski' vacua (the in/out shockwave geometry vacua) are not unitarily equivalent with respect to the full BMS group: schematically, we have $\ket{0'}\coloneqq \ket{0,\alpha'}\neq\ket{0,\alpha} \equiv  \ket{0}$.
}

We time-evolve the initial state $\rho_0$ perturbatively by performing the Dyson series expansion of the time-evolution operator to second order in $\lambda$,
\begin{subequations}
    \begin{align}
    U &= \openone + U^{(1)} + U^{(2)} +O(\lambda^3)\,,\\
    U^{(1)} &=  -i\int_{-\infty}^\infty d\tau\,H_{I}(\tau)\,,\\
    U^{(2)} &= -\int_{-\infty}^\infty dt\int_{-\infty}^\infty dt'\,\Theta(\tau-\tau') H_{I}(\tau)H_{I}(\tau')\,.
    \end{align}
    \label{eq: dyson}
\end{subequations}
\noindent Tracing out the field degrees of freedom gives the reduced density matrix of the qubit
\begin{align}
    \rho_{d} = \Tr_\phi\rr{U\rho_0 U^\dagger} = \rho_0+\rho^{(1)}+\rho^{(2)}+ O(\lambda^3)\,,
\end{align}
where $\rho^{(k)}$ is the correction of order $\lambda^k$:
\begin{subequations}
\begin{align}
    \rho^{(1)} &= \Tr_\phi \rr{\rho_0U^{(1)\dagger}} +  \Tr_\phi \rr{U^{(1)} \rho_0}\,,\\
    \rho^{(2)} &= \Tr_\phi \Bigl(\rho_0U^{(2)\dagger}\Bigr) +\Tr_\phi \rr{U^{(2)} \rho_0}+\Tr_\phi \rr{U^{(1)} \rho_0U^{(1)\dagger}}\,.
\end{align}
\end{subequations}
Since the field is initially in the vacuum state, $\rho^{(1)}=0$ due to the vanishing of vacuum one-point functions (in fact $\rho^{(2k+1)}=0$ for non-negative integer $k$). The leading order contribution is therefore of order $\lambda^2$, which in the ordered basis $\{\ket{g},\ket{e}\}$ reads
\begin{equation}
    \rho_d=\begin{pmatrix}
    1-P& 0\\
    0 & P
    \end{pmatrix}+O(\lambda^4)\,,
    \label{eq: reduced-detector-state-one}
\end{equation}
where
\begin{equation}
    P = \lambda^2\int d\tau\,d\tau'\,\chi(\tau)\chi(\tau') e^{-i\Omega(\tau-\tau')}W(\sx[\tau],\sx[\tau'])\,,
    \label{eq: transition}
\end{equation}
and $W(\sx[\tau],\sx[\tau'])$ is the pullback of the Wightman function along the detector's trajectory. Here, $P$ can be understood as the detector transition probability from the ground state to the excited state.

\subsection{Entanglement harvesting protocol}

Implementing the entanglement havesting protocol requires generalizing to two pointlike UDW detectors, denoted $A$ and $B$, which have proper times $\tau_j$ and spacetime trajectories $\sx_j[\tau_j],\, j = A,B$.  In general, the detectors may have different coupling constants $\lambda_j$, switching functions $\chi_j(\tau_j)$, and energy gaps $\Omega_j$.  However, in what follows, we consider a simplified scenario with identical detectors such that $\lambda_j=\lambda$ and $\Omega_j=\Omega$. As in the single detector case, we will also assume that the switching functions are Gaussian functions peaked about $\tau_{j,0}$ with Gaussian width $T$.

Since the spacetime has a global coordinate time $t$, it is convenient to use this to define the full interaction Hamiltonian of the detector-field system:
\begin{align}
    H^t_I(t) &= H^t_{I,A}(t)\otimes \openone_B + \openone_A \otimes H^t_{I,B}(t) \notag\\
    &=  \frac{d\tau_A(t)}{dt} H^{\tau_A}_{I,A}(\tau_A(t))\otimes \openone_B + \openone_A \otimes \frac{d\tau_B(t)}{dt} H^{\tau_B}_{I,B}(\tau_B(t))\,,
\end{align}
where $H^t_I$ generates time translation with respect to $t$ (similarly for $H^{\tau_j}_I$). The second equality follows from time-reparametrization invariance of the Hamiltonian \cite{Tales2020GRQO} which allows us to evolve the system with respect to this one time $t$.

The calculation involved in the entanglement harvesting protocol proceeds now similarly to the previous section. First, we prepare the joint detector-field system in the initially-uncorrelated state
\begin{align}
    \rho_0 &= \ket{g}_A\!\bra{g}_A\otimes  \ket{g}_B\!\bra{g}_B\otimes\ket{0}\!\bra{0}\,,
\end{align}
where we take $\ket{0}$ to be the `in' vacuum state. Evolving the state perturbatively using the Dyson series \eqref{eq: dyson} and tracing out the field degrees of freedom gives the reduced density matrix of the bipartite qubits $\rho_{AB}$.  As before, the leading order contribution is of order $\lambda^2$, which in the ordered basis $\{\ket{g}_A\ket{g}_B,\ket{e}_A\ket{g}_B,\ket{g}_A\ket{e}_B,\ket{e}_A\ket{e}_B\}$ reads
\begin{align}
    \rho_{AB} =
    \begin{pmatrix}
    1-P_A-P_B & 0 & 0 & \mathcal{M}^* \\ 0 & P_B & C^* & 0 \\
    0 & C & P_A & 0\\
    \mathcal{M} & 0 & 0 & 0
    \end{pmatrix}+O(\lambda^4)\,,
    \label{eq: reduced-detector-state-two}
\end{align}
where 
\begin{align}
    C &= \lambda^2\int d\tau_A\,d\tau_B\,\chi_A(\tau_A)\chi_B(\tau_B) e^{-i\Omega(\tau_A-\tau_B)}W(\sx_A[\tau_A],\sx_B[\tau_B])\,,\\
    \mathcal{M} &= -\lambda^2\int d\tau_A\,d\tau_B\,\chi_A(\tau_A)\chi_B(\tau_B) e^{i\Omega(\tau_A+\tau_B)}\Theta[t(\tau_A)-t(\tau_B)]W(\sx_A[\tau_A],\sx_B[\tau_B])\notag\\
    &\phantom{=}-\lambda^2\int d\tau_A\,d\tau_B\,\chi_B(\tau_B)\chi_A(\tau_A) e^{i\Omega(\tau_A+\tau_B)}\Theta[t(\tau_B)-t(\tau_A)]W(\sx_B[\tau_B],\sx_A[\tau_A])\,,
\end{align}
and $P_A$ and $P_B$ are the transition probabilities of each detector \eqref{eq: transition}.

Entanglement can be harvested because  $\rho_{AB}$ will be an entangled state for suitable range of detector parameters. This can be verified using computable measures of entanglement such as the \textit{negativity} or \textit{concurrence} \cite{Wotters1998entanglementmeasure,Horodecki996separable,Vidal2002negativity}.
For the purpose of this paper, we concentrate on the concurrence which (for the joint density matrix of the detectors \eqref{eq: reduced-detector-state-two}) is given by~\cite{smith2016topology}
\begin{equation}
 \mathcal{C}[\rho_{AB}]=   
 2\max\{0, |\mathcal{M}|-\sqrt{P_A P_B}\} +O(\lambda^4)\,.
\end{equation}

In this form, we see that entanglement between the two detectors admits a simple interpretation, namely it is a competition between non-local quantum correlations coming from $\mathcal{M}$ and the noise terms $\sqrt{P_AP_B}$ coming from each detector's excitations as they interact locally with the field.

\begin{figure}[h!]
\centering
\includegraphics[width=\textwidth]{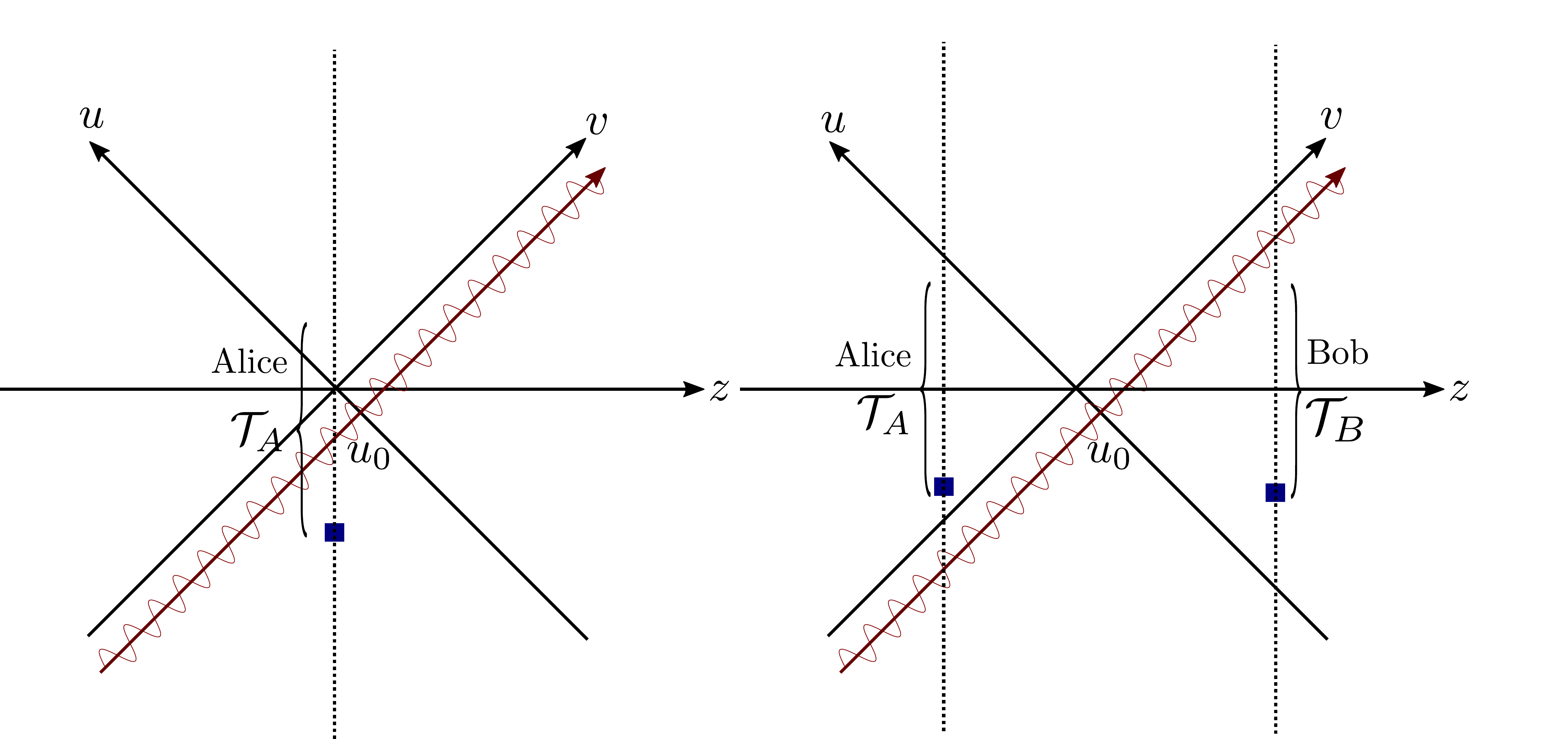}
\caption{{\bf Single detector and entanglement harvesting protocol.} {\em Left:} Alice and her detector cross the shockwave whilst interacting with the field for time ${\cal T}_A$. (For example,  ${\cal T}_A$ could correspond to the interval $[-4T+\tau_0,4T+\tau_0]$ around the peak of the Gaussian switching \eqref{eq:Gauss_Switch}.) {\em Right.} Alice and Bob are located on either side of the shockwave and never cross the shockwave while interacting with the field for a duration ${\cal T}_{A/B}$. This interaction leads to entanglement between Alice and Bob.\label{fig:UdWSetup}}
\end{figure}

\subsection{Detectors' worldlines}

Here we are going to analyze the detector's response and entanglement harvesting in the shockwave spacetime, specializing to the physical case $D=4$. For this we need to specify the trajectories of the detectors in this  spacetime and calculate the corresponding pullbacks of the Wightman functions. However, there is a complication in that the presence of the shock alters the geodesics---in general both spacelike and timelike geodesics will suffer from discontinuities. Since our calculation involves comparing entanglement harvesting in the shockwave geometry with that in Minkowski space for the special case when the detectors are static at constant $x^j$, we need to find comparable trajectories.

As shown in the Appendix~\ref{app:B}, it turns out that in the pointlike regime and for the special case of the planar shockwave with the `Gaussian' form factor \eqref{eq:Gauss_profile}, there is a special choice of the impact parameter $\vec{b}=0$ (in the given coordinates) so that the geodesics on the corresponding  codimension-2 $(t,z)$-plane are identical to those in Minkowski space. This yields the two natural scenarios for our calculations displayed in Fig.~\ref{fig:UdWSetup}. Namely, to discuss the response of a single detector to the shockwave, we consider 
a static detector, placed at 
\begin{equation}
\sx(\tau) = (t(\tau), z(\tau), \vec{x}(\tau))=(\tau, 0, \vec{0})\,,    
\end{equation}
which remains geodesic even when hit by the shockwave, as displayed on the left in Fig.~\ref{fig:UdWSetup}. For the entanglement harvesting, we shall consider two (geodesic) static detectors\footnote{In our setup, we say that we have arranged two detectors so that neither of them crosses the shock if the Gaussian tails of the switching functions crossing the shock are negligible. In practice, we truncate the exponential tails of the switching and declare that the detectors do not touch the shock if the probabilities are equal to that in Minkowski space up to the fifth significant figure.}, 
\begin{align}
    \sx_A(\tau_A) &= (t(\tau_A), z(\tau_A), \vec{x}(\tau_A))=(\tau_A, z_A, \vec{0})\,, \label{eq: Worldline_A}\\  
    \sx_B(\tau_B) &= (t(\tau_A), z(\tau_A), \vec{x}(\tau_B))=(\tau_B, z_B, \vec{0})\,, \label{eq: Worldline_B}
\end{align}
as displayed on the right in Fig.~\ref{fig:UdWSetup}. The important feature of this choice is that the  proper spatial distance between these detectors, $L = |z_B-z_A|$, remains the same as in the Minkowski space, since the constant-$t$ slices restricted to $\vec{x}=0$ are not affected by the shock. It is worth mentioning that in our setup the two detectors are placed in the longitudinal direction rather than the transverse direction, as was done for the linearized gravity case in \cite{Smith2020harvestingGW}.

\section{Results}\label{sec:results}

\subsection{Single detector's response to a shockwave}
Before we turn to entanglement harvesting we will discuss the response of a single detector to the shockwave. Previous studies~\cite{Compere:2019rof,Majhi2020shock} have suggested that an observer who probes a scalar field in the ground state will only detect the Minkowski results. 
Moreover, in~\cite{Smith2020harvestingGW} it was shown that a single stationary UDW detector will not detect a linearized plane gravitational wave. These results seem intuitive on the grounds that a single test particle (classical observer) cannot reveal the presence/absence of gravitational waves.

On the other hand, it is clear from the Wightman function~\eqref{eq:finalW} that, despite the local flatness, a single detector should be able to detect the gravitational shockwave as it passes by. Indeed, if we allow the detector's switching function $\chi(\tau)$ to cross the shockwave, then the detector response will \emph{not} reduce to the flat space one. Ultimately, this is because we are integrating over the Wightman function where the shockwave form factor does not cancel (essentially $\Delta\Theta_u\neq 0$). 

\begin{figure}[ht]
    \centering
    \includegraphics[scale=1.2]{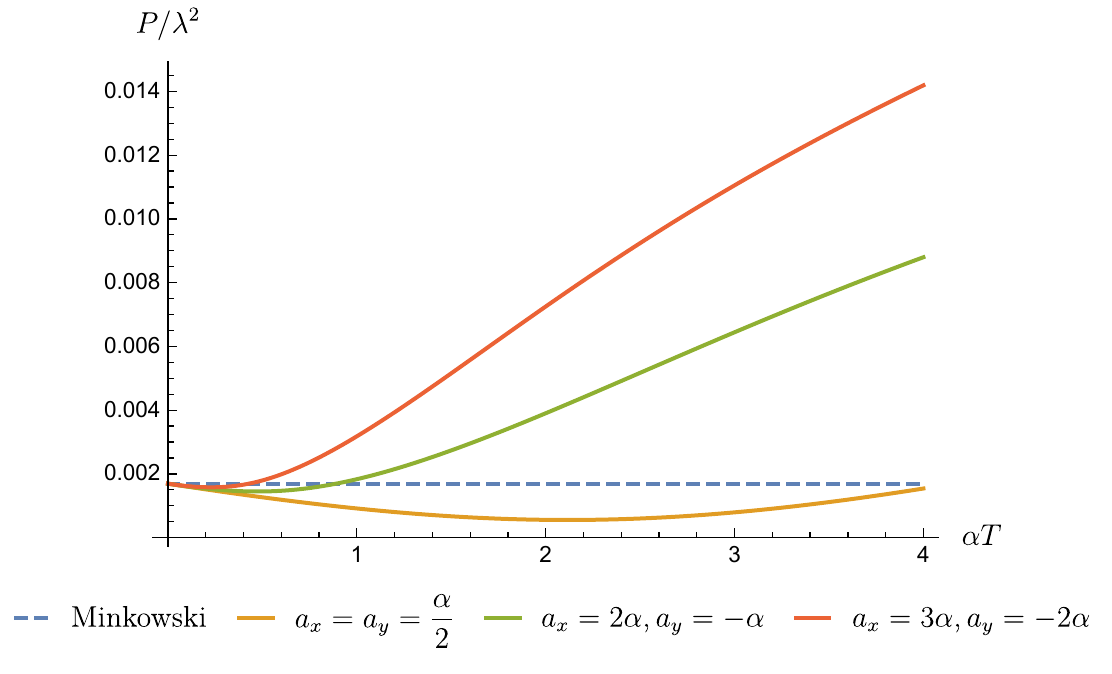}
    \caption{{\bf Excitation probability of a single detector.} 
    We display the detector transition probability for three shockwave profiles with the same energy density $\alpha$: $a_x=2\alpha, a_y=-\alpha$ (red curve), $a_x=3\alpha, a_y=-2\alpha$ (green curve), $a_x=a_y=\frac{\alpha}{2}$ (orange curve) and we compare them to the Minkowski response (dashed blue curve). The scenario follows Left Fig.~\ref{fig:UdWSetup}: the detector is located at the origin, with the switching peak at $t=0$, and energy gap $\Omega T = 2$, the shockwave is localized at $u=0$. Note that, in this parametrization, when $\alpha=0$, $f(\vec{x})=0$ identically and we recover Minkowski space.}
    \label{fig: one-detector-trace}
\end{figure}

To demonstrate this effect, we plot the excitation probability of the detector as a function of the strength of the planar shockwave in Fig.~\ref{fig: one-detector-trace}. This is characterized by various eigenvalues of the matrix $A$, which we parameterize in terms of the wave energy density 
\begin{equation}
\alpha=\Tr(A)=8\pi G \rho\,.    
\end{equation} 
The response for Minkowski space is displayed by the blue dashed curve and compared to three wave profiles depicted by solid orange, green, and red curves.  
Interestingly, for small enough $\alpha$ the excitation probability of the detector actually decreases compared to the flat space result. However, increasing the strength/energy density of the shockwave ultimately leads to a larger excitation probability. In particular, this means that for a given profile of the wave, there exists an amplitude for which a single detector cannot distinguish the wave from the flat space. Let us also note that the symmetric shockwave  ($a_x=a_y=\alpha/2$) has a much smaller effect than the other cases. The reason for this can be understood as follows:
if one holds the energy density, $\alpha$ of the shockwave constant, then the deviation from the Minkowski Wightman function in $\eqref{eq: Wightman-Gauss}$ is (when $\vec{x}=0=\vec{X}$) minimized for $a_x=a_y$. 

To summarize, we have clearly illustrated that the gravitational shockwave passing through the UDW detector leaves an observable, at least in principle, imprint on the detector.

\subsection{Comparison to linearized gravitational waves}

Our results should be contrasted with the conclusions found in the context of linearized gravity: how did we avoid the conclusions of \cite{Smith2020harvestingGW}? To begin answering this question, we recall that the metric of the linearized plane gravitational wave considered in \cite{Smith2020harvestingGW} reads
\begin{align}
    d s^2 = -d u d v+(1+B\cos\omega u)d x^2+(1-B\cos\omega u)d y^2\,,
\end{align}
where $B$ is the wave amplitude and $\omega$ its frequency.\footnote{Note that the employed coordinates are not the Brinkmann coordinates we used in \eqref{eq:metric}, rather, they are a special case of the so called BJR coordinates \cite{zhang2017memory}.}
This metric is a solution to the vacuum linearized Einstein equations, that is, it is only a solution to the full non-linear vacuum equations to order $O(B)$. Furthermore, the authors showed that the first-order correction $\Delta_1W_{\text{GW}}(\sx,\sx')$ to the Wightman function restricted to the trajectory $\Delta x=\Delta y =0$ vanishes, hence the probability is not affected by the gravitational wave at order $O(B)$. 

We can in fact show that for this metric there is a closed-form expression for the Wightman function\footnote{In fact, the integral of (A3) in~\cite{Smith2020harvestingGW} is just Gaussian and so the techniques of Appendix \ref{app:A} in this paper can straightforwardly be applied.} such that the $O(B^2)$ correction to the Minkowski Wightman function $\Delta_2 W_{\text{GW}}(\sx,\sx')$ for $\Delta x=\Delta y=0$ does not vanish. It is given by
\begin{align}
    \Delta_2 W_{\text{GW}}(\sx,\sx')=-B^2\frac{ \omega^2 (\Delta u-i\epsilon)^2 [\cos^{2} (\omega u )+\cos^{2} (\omega u')]-2[\sin (\omega u)-\sin (\omega u')]^2}{16 \pi ^2 \omega^2 (\Delta u-i\epsilon)^3 (\Delta v-i\epsilon)}\,.
\end{align}
While the second-order correction cannot be used on the grounds that it goes beyond the given (linear)  approximation, it suggests that a full non-linear gravitational wave solution to the Einstein equations could in general have nontrivial signature to the detector response. This is exactly the case for the `Gaussian' shockwave considered in this paper with the Wightman function \eqref{eq: Wightman-Gauss}.

To see how this arises, let us consider an expansion in small shockwave amplitude, that is, $||AT||\ll 1$ (using switching duration $T$ as a reference scale). Specifying to $\Delta x=\Delta y=0$, the general expression \eqref{eq: Wightman-Gauss} yields the following expansion in $A$: 
\begin{align}\label{eq:Wight_pert}
    W(\sx,\sx') &= -\frac{1}{4 \pi ^2 (\Delta u-i\epsilon) (\Delta v-i\epsilon)}\left(1-\frac{\Tr(A)\Delta\Theta_u (u-u_0-i\epsilon) (u'-u_0+i\epsilon) }{2 (\Delta u-i\epsilon)}\right.\nonumber\\
    &\left.+\frac{(2\Tr(A^2)+\Tr(A)^2) [\Delta\Theta_u (u-u_0-i\epsilon) (u'-u_0+i\epsilon) ]^2}{8 (\Delta u-i\epsilon)^2}+O(A^3) \right)\,,
\end{align}
which has a first-order correction to the flat-space Wightman function proportional to the trace of $A$. This result also explains why a slight increase in the strength of the shock can actually decrease the transition probability---the first order correction is negative.

\begin{figure}[h!]
    \centering
    \includegraphics[scale=1.2]{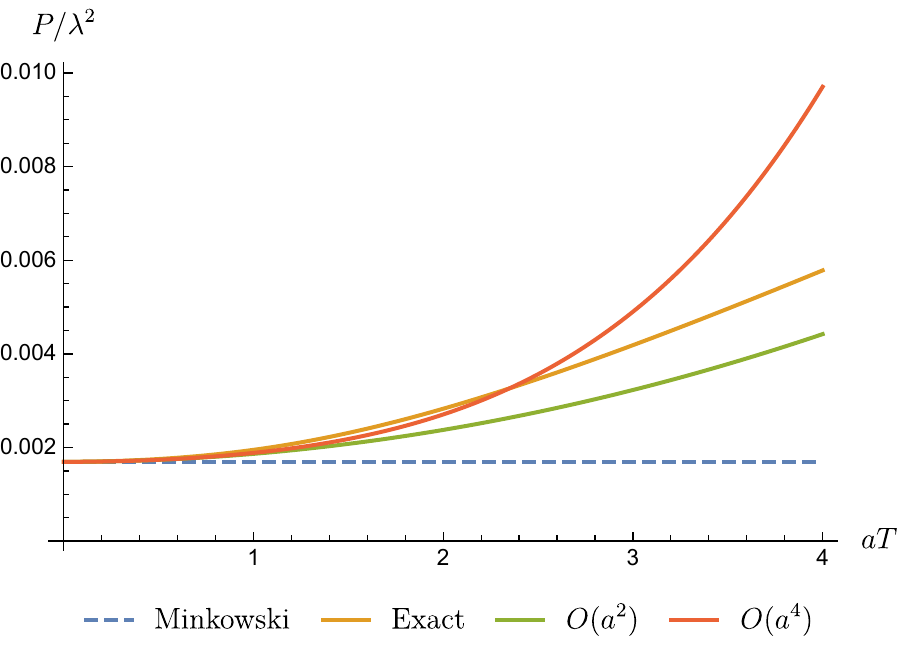}
    \caption{{{\bf `{GW} amplitude expansion' 
    of the excitation probability for source-less waves.} Considering the source-less shockwaves with $a_x=-a_y=a$, we display the exact (orange curve), the $O(a^2)$ (green curve), and the $O(\alpha^4)$ (red curve) amplitude expansions of the transition probability and compare them to that of the Minkowski space (blue dashed curve). As previously, the detector is located at the origin, with the switching peak at $t=0$, the shockwave follows $u=0$, and $\Omega T = 2$.} 
    }
    \label{fig: one-detector-traceless} 
\end{figure}

Suppose now that we want to mimic the setup of the vacuum plane gravitational wave in $D=4$, obtained via linearized gravity. For this we choose a sourceless, but \emph{non-trivial}, shockwave; $\alpha=\tr(A)=0$ with non-vanishing 
\begin{equation}
    a_1=-a_2=a\,,\quad aT\ll 1\,.    
\end{equation}
Then, as in~\cite{Smith2020harvestingGW}, we see that the contribution of the shockwave comes in only at second-order. Furthermore, it can be checked that all corrections with odd powers of $a$ vanish so that the subleading corrections are of $O(a^{2n})$. We plot the excitation probability as a function of the shock profile diagonal element $a$  in Fig.~\ref{fig: one-detector-traceless} up to fourth order in $a$. 

This result shows that if we regard $a$ as taking the role analogous to gravitational wave amplitude, then we obtain `similar results' to \cite{Smith2020harvestingGW}, in that, at linear order $O(a)$ the gravitational (shock)wave does not modify the excitation probability of a single detector restricted to a static trajectory. On the other hand, since the shockwave is a solution to the full non-linear Einstein equations, higher order corrections always exist and these allow a single static detector to see the shockwave as it passes by. We expect this to be a generic feature of non-linear gravitational waves.

\subsection{Entanglement harvesting}

The single detector result shows that if we allow the shockwave to pass through the {(switched on)} detector, then there is a clear signature that changes the excitation probability of the detector. This is the case even if the shockwave is sourceless. A natural question that arises is whether two detectors, localized on two different sides of the shock (and thence in the flat space for the entire duration of the measurement), 
are still able to detect the presence of the shockwave. 
The entanglement harvesting protocol is well-suited to address this type of question: the spirit is similar to \cite{VerSteeg2009entangling}, where it was shown that two inertial detectors can distinguish the QFT vacuum on de Sitter background versus the thermal state in Minkowski space, even though a single detector will register an identical response in both cases.

\begin{figure}[tp]
    \centering
    \includegraphics[scale=1.2]{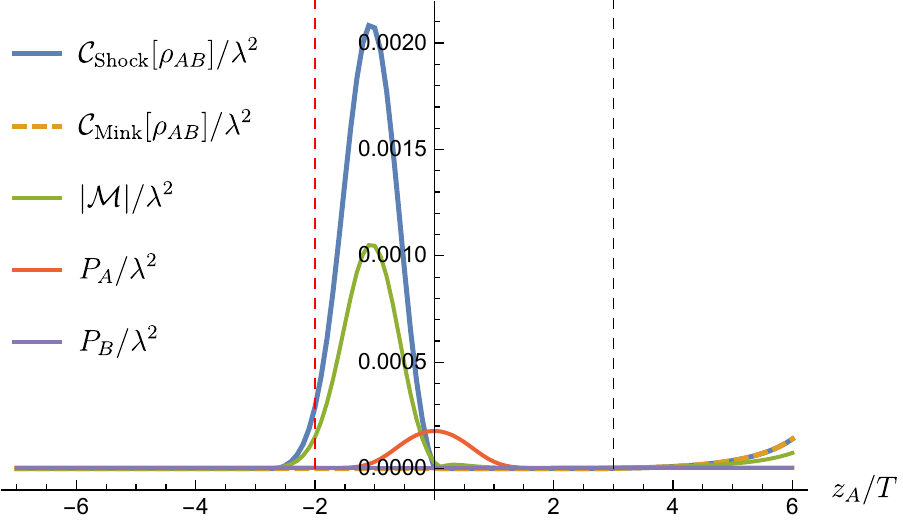}
    \vspace{.2cm}
    \includegraphics[scale=1.2]{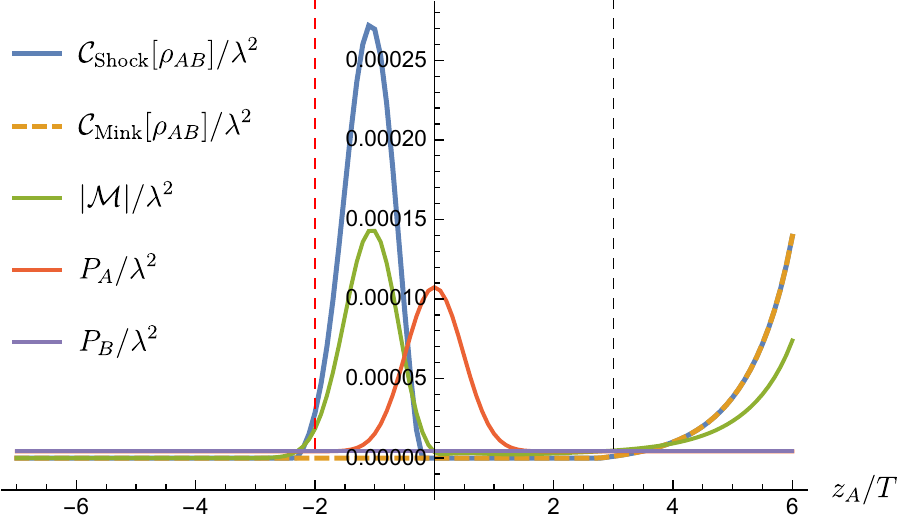}
    \caption{\textbf{Concurrence, non-local correlation and local noise contributions for two detectors.} The plots show the concurrence between Alice and Bob's detectors and the corresponding local/non-local contributions as a function of Alice's position. We show the concurrence for Minkowski space (orange dashed curve) as a reference. For both plots, the shockwave is localized at $u=0$, Bob is fixed at $z_B = 7T$, and Alice's position $z_A$ is varied.  We set $\Omega T = 3.75$. \textit{Top:} Shockwave with nonzero trace, i.e, $a_xT = a_yT=1$. \textit{Bottom:} Traceless shockwave $a_xT = -a_yT=1$. In each case the vertical dashed line in red marks the position of Alice's detector where the shock begins to alter the excitation probability $P_A$, thus Alice has not crossed the shock for $z_A\lesssim -2T$. The vertical dashed line in black marks the position of Alice's detector when the (effectively compact) support of Gaussian switching begins to cross causal future/past of Bob's detector, thus allowing both detectors to communicate via the quantum field. This manifests as increase in concurrence for $z_A\gtrsim 3T$.}
    \label{fig: concurrence-shockwave}
\end{figure}

In Fig.~\ref{fig: concurrence-shockwave} we show that the entanglement harvesting protocol indeed has a specific quantum signature in terms of concurrence. Namely, therein, we plot the concurrence $\mathcal{C}[\rho_{AB}]$ for two detectors $A$ and $B$ associated with two static observers Alice and Bob along the worldlines in \eqref{eq: Worldline_A} and \eqref{eq: Worldline_B}. We also display the corresponding nonlocal $|{\cal M}|$ term and the local noise contributions (given by the probabilities of each detector, $P_A$ and $P_B$). The top figure corresponds to a shockwave with trace $a_xT = a_yT=1$ and the bottom one is traceless $a_xT = -a_yT=1$.

In Fig.~\ref{fig: concurrence-shockwave}, Bob is fixed at constant $z_B=7T$ located on the $u<u_0=0$ part of the shockwave geometry. We study how concurrence varies when Alice is placed on different static trajectories $z_A$, starting from $z_A=-7T$ in the $u>u_0=0$ region. We also include the concurrence for the same setup in full Minkowski spacetime as a reference. Note that, when compared to Figs.~\ref{fig: one-detector-trace} and \ref{fig: one-detector-traceless}, we have considered here a much bigger detector energy gap, setting it to $\Omega T=3.75$. This is because, in general, the excitation probability is highly suppressed with higher $\Omega $ (physically it is harder to excite with larger energy gap), and $\Omega$ needs to be sufficiently large for the non-local contribution $|\mathcal{M}|$ to eventually dominate over local noise $\sqrt{P_AP_B}$. Consequently, the noise contribution for $\Omega T = 3.75$ is an order of magnitude smaller than in the previous figures where $\Omega T=2$.

Fig.~\ref{fig: concurrence-shockwave} shows that when the shock passes through Alice's detector, the concurrence is greatly amplified relative to the Minkowski equivalent, with a larger amount of entanglement extracted for the shock with nonzero stress-energy. Therefore, the shockwave increases the non-local quantum correlations encoded in $|\mathcal{M}|$ more than the local excitation encoded in the noise term $\sqrt{P_AP_B}$ (effectively $P_A$ since we hold $P_B$ fixed). However, notice that the concurrence is amplified even \textit{before} Alice reaches $z_A\approx -2T$ (shown as vertical dashed line) where $P_A$ starts to differ from the Minkowski value. In other words, there is a small window $z_A\lesssim -2T$ where the detector responses are indistinguishable from the Minkowski value but, nonetheless, the two detectors extract greater quantum correlations than in true Minkowski space. Therefore, two spacelike separated detectors\footnote{The vertical black dashed line in Fig.~\ref{fig: concurrence-shockwave} indicates where Alice and Bob become causally connected.} also can detect quantum imprint of gravitational shockwave even though locally each detector perceives the background geometry to be true Minkowski space.

\begin{figure}[th]
    \centering
    \includegraphics[scale=1]{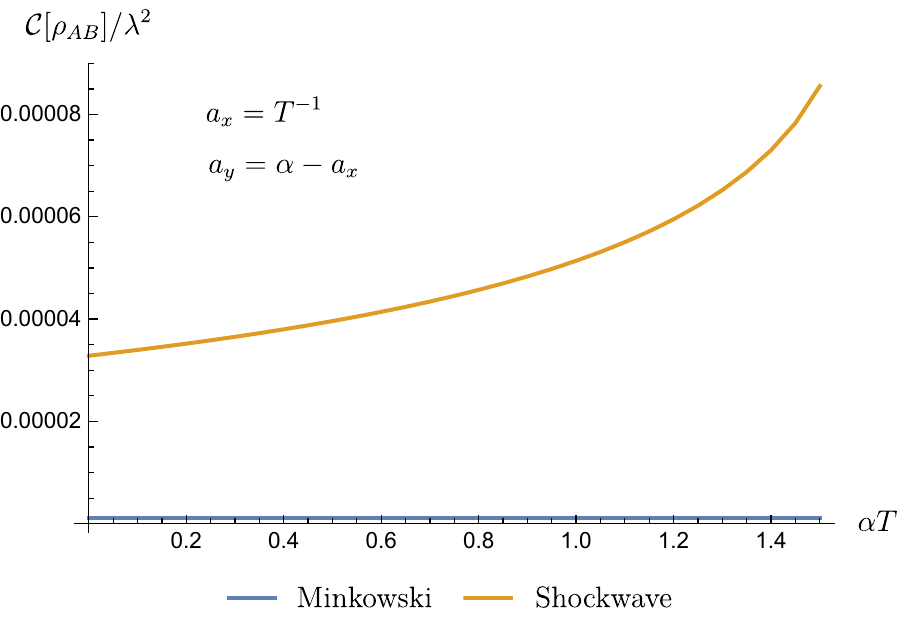}
    \caption{\textbf{Concurrence as a function of $\alpha=\Tr(A)$}. By fixing $a_xT$ (here we set $a_xT=1$ for concreteness), the trace characterizes the relative magnitude of $a_y$. Here we set $\Omega T = 3.75$, and have the detectors fixed  at $z_A = -2T$ and $ z_B = 2T$. Note that from Fig.~\ref{fig: concurrence-shockwave} we know that the detectors do not encounter the shockwave during their switched-on period. 
    \label{fig:TurningSource} }
\end{figure}

It is worth checking how the concurrence varies with the shock profile and hence the stress-energy tensor. For concreteness, we fix one component of the matrix $A=\text{diag}(a_x,a_y)$ to some value $a_x T=1$ and express $a_y = \alpha-a_x$ where as before $\alpha=\Tr(A)$, and plot this setup in Fig.~\ref{fig:TurningSource}. The two detectors are fixed at $x_A=-2T$ and $x_B=2T$, thus they have not yet crossed the shock at $u=u_0=0$ (as can be checked from Fig.~\ref{fig: concurrence-shockwave}). We include the Minkowski value (in blue) as a baseline. We see that first of all the traceless case, which corresponds to sourceless shockwave, has lower concurrence than the case with nonzero trace (nonzero stress-energy). This result is consistent with what we found in Fig.~\ref{fig: concurrence-shockwave}. We also see that the amount of entanglement extracted by the two detectors is able to probe a constant DC shift (reminiscent of the clasical memory effect) due to the step function $\Theta(u-u_0)$ induced by the shockwave, which is not observable for a single detector localized away from the shockwave (since $\Delta\Theta_u=0$).

\section{Conclusions}
\label{sec:conclusions}

We have considered local observers carrying quantum-mechanical detectors interacting with quantized scalar fields on a gravitational shockwave background geometry. Our work was motivated by a series of studies~\cite{Kolekar2018quantummemory,Compere:2019rof,Majhi2020shock,Ferreira:2020whz} which probed the 
effect of these shockwaves on the global states and properties of the quantum field. In this context, it was observed that a matter-induced supertranslation due to a propagating shockwave only has physical consequences on non-vacuum states of the field, and that the spectrum of Hawking/Unruh radiation remains unaffected~\cite{Compere:2019rof,Majhi2020shock,Ferreira:2020whz}. These results complimented the notion of `soft hairs' encoded in BMS supertranslations and their connection with the infrared triangle~\cite{Hawking:2016sgy,strominger2018lectures}.

Contrary to \cite{Compere:2019rof,Majhi2020shock,Ferreira:2020whz}, we have demonstrated that local observers interacting with the scalar field vacuum via Unruh--DeWitt detectors \emph{can} see non-trivial results. A single detector has a non-trivial probability of getting excited from a gravitational shockwave passing through its worldline, in contrast to the linearized gravity results presented in \cite{Smith2020harvestingGW}. {We have also shown} that if Alice and Bob, each carrying a UDW detector, are placed on different sides of the shockwave, {they can use the  entanglement} harvesting protocol to discover the non-trivial impact of the shockwave on the scalar field vacuum. This happens even when their detectors do not cross the shockwave and remain in flat space for the entire duration for which they are switched on.

So why does Alice see the shockwave? As mentioned above, in the first work~\cite{Compere:2019rof} only the globally defined number operators and Bogoliubov coeffecients were considered. These global features do not necessarily capture all the details of a local finite time interaction with the field~\cite{Garay2016anti-unruh,Brenna2016anti-unruh,Henderson2020anti-hawking,Dappiaggi2021anti-hawking}. In particular, more physical interactions with the field can lead to unintuitive results like the anti-Hawking/Unruh effect~\cite{Garay2016anti-unruh,Brenna2016anti-unruh,Henderson2020anti-hawking,Dappiaggi2021anti-hawking} and the `standard' results are only recovered by letting the interaction happen over a long time scale. Moreover, the second work~\cite{Majhi2020shock} is only concerned with an observer completely localized on the Minkowski part in either side of the shockwave. Since the spacetime is flat on either side, the Wightman function is necessarily equal to the Wightman function in the standard Poincar\'e-invariant vacuum in Minkowski space. This emphasizes that a concrete notion of local observers can be important in the context of extracting information from quantum fields.

Finally our results provide another situation where certain gravitational wave effects manifest at the level of Wightman functions and hence affect localized interactions with an external probe. In contrast to linearized gravitational waves~\cite{Smith2020harvestingGW} we have seen that a single detector is sensitive to a shockwave. This is a manifestation of the nonlinear characteristics of gravity and we expect it to be a generic feature of quantum gravitational wave detection.

Thus, local detection and entanglement harvesting can physically extract quantum imprints of gravitational shockwaves.

\section*{Acknowledgment}
We thank Ma\"it\'e Dupuis for organising the PSI Winterschool wherein this project was initiated and our minds and bodies stretched.
F.G. is funded from the Natural Sciences and Engineering Research Council of Canada (NSERC) via a Vanier Canada Graduate Scholarship. 
E.T. acknowledges support from Mike--Ophelia Lazaridis Fellowship.
This work was also partially supported by NSERC and partially by the Perimeter Institute for Theoretical Physics. Research at Perimeter Institute is supported in part by the Government of Canada through the Department of Innovation, Science and Economic Development Canada and by the Province of Ontario through the Ministry of Colleges and Universities. 
Perimeter Institute and the University of Waterloo are situated on the Haldimand Tract, land that was promised to the Haudenosaunee of the Six Nations of the Grand River, and is within the territory of the Neutral, Anishnawbe, and Haudenosaunee peoples.

\appendix

\section{Wightman in gravitational shockwave spacetimes}
\label{app:A}

In this appendix we present the explicit calculation of some of the integrals necessary to arrive at the forms of the Wightman function in the main text \eqref{eq:finalW} and \eqref{eq: Wightman-Gauss}. 

\subsection{From Gaussian integrals to the Wightman function}

We begin by recalling the familiar Gaussian integral: provided $\Im{a}>0$,
\begin{equation}\label{eq:Gauss_Int}
    \int_{-\infty}^{+\infty} \,dx\, e^{iax^2+ibx}=\sqrt{\frac{i\pi}{a}}e^{-i\frac{b^2}{4a}}\,.
\end{equation}
This result directly applies to the mode functions
\begin{align}
         \phi^\tin_{k_-,\vec{k}}=N_{k_-}e^{-ik_-v}e^{i\vec{k}\cdot\vec{x}}\int \frac{d\vec{x'}d\vec{k'}}{(2\pi)^{D-2}}e^{i(\vec{k}-\vec{k'})(\vec{x'}-\vec{x})-\frac{i\vec{k'}^2}{4k_-}(u-u_0)+ik_-\Theta(u-u_0)f(x')}\,,
\end{align}
 where we can perform the integral over $\vec{k}'$ using \eqref{eq:Gauss_Int} by adding an infinitesimal $i\epsilon$ to avoid branch cuts
\begin{align}
    \int d\vec{k}'\, e^{i\vec{k}'\cdot (\vec{x}-\vec{x'})}e^{-i\frac{|\vec k'|^2}{4k_-}(u-u_0-i\epsilon)} &= \left(\frac{-4\pi i k_-}{u-u_0-i\epsilon}\right)^{\frac{D-2}{2}}\exp\left[\frac{ik_-(\vec{x}-\vec{x}')^2}{u-u_0-i\epsilon}\right]\,.
\end{align}
Thus 
\begin{multline}
    \phi^\tin_{k_-,\vec{k}}=\left(\frac{-4\pi i k_-}{u-u_0-i\epsilon}\right)^{\frac{D-2}{2}} N_{k_-}e^{-ik_-v}e^{i\vec{k}\cdot\vec{x}}\\
    \times\int \frac{d\vec{x'}}{(2\pi)^{D-2}} \exp\left[\frac{ik_-(\vec{x}-\vec{x}')^2}{u-u_0-i\epsilon}+ik_-\Theta(u-u_0)f(x')\right]\,.
\end{multline}
Plugging this expression into the mode sum version of the Wightman function
\begin{equation}
W(\sx,\sy)=\int dk_-d\vec{k} \,\phi^\tin_{k_{-},\vec{k}}(\sx)\,\phi^{\tin\,*}_{k_{-},\vec{k}}(\sy)\,,
\end{equation}
and integrating over the $\vec{k}$ directions we find \eqref{eq:WStart} of the main text:
\begin{align}
    W(u,v,\vec{x};U,V,\vec{X})&= \int \frac{dk_{-}}{2(2\pi)^{D-1}}\,e^{-ik_{-}(\Delta v-i\epsilon)}\int\frac{d\vec{x}'}{(2\pi)^{D-2}}\,e^{ik_{-}\left[\Theta_uf(\vec{x}')-\Theta_Uf(\vec{x}')\right]}\notag\\
    &\times\frac{(4\pi )^{D-2}k_-^{D-3}}{[(u-u_0-i\epsilon)(U-u_0+i\epsilon)]^{\frac{D-2}{2}}}\exp\left[\frac{ik_-(\vec{x}-\vec{x}')^2}{u-u_0-i\epsilon}-\frac{ik_-(\vec{X}-\vec{x}')^2}{U-u_0+i\epsilon}\right]\,.
\end{align}
Here we have introduced $\Delta v=v-V$ and $\Delta\Theta_u=\Theta(u-u_0)-\Theta(U-u_0)$. Next we integrate over $k_-$:
\begin{align}
    &\int_0^\infty dk_-\,k_-^{D-3}\exp \left(-ik_-\left[\Delta v-i\epsilon-f(\vec{x}')\Delta\Theta_u\right]\right)\exp\left[\frac{ik_-(\vec{x}-\vec{x}')^2}{u-u_0-i\epsilon}-\frac{ik_-(\vec{X}-\vec{x}')^2}{U-u_0+i\epsilon}\right]\notag\\
    &= \Gamma (D-2)(i)^{2-D} \left(\left[\Delta v-i\epsilon-f(\vec{x}') \Delta\Theta_u\right]- \frac{(\vec{x}-\vec{x}')^2}{u-u_0-i\epsilon}+\frac{(\vec{X}-\vec{x}')^2}{U-u_0+i\epsilon}\right)^{2-D}\,,
\end{align}
and thence recover \eqref{eq:finalW}
\begin{align}
    W(u,v,\vec{x};U,V,\vec{X})&= \frac{\Gamma (D-2)(i)^{2-D}}{2(2\pi)^{D-1}}\frac{(4\pi)^{D-2}}{[(u-u_0-i\epsilon)(U-u_0+i\epsilon)]^{\frac{D-2}{2}}} \int\frac{d\vec{x}'}{(2\pi)^{D-2}}\notag\\
    &\hspace{0.5cm}\times \left(\left[\Delta v-i\epsilon-f(\vec{x}') \Delta\Theta_u\right]- \frac{(\vec{x}-\vec{x}')^2}{u-u_0-i\epsilon}+\frac{(\vec{X}-\vec{x}')^2}{U-u_0+i\epsilon}\right)^{2-D}\,,\label{eq:finalW_app}
\end{align}
which appeared in the main text. This is as far as we can go without specifying $f(\vec{x})$.

To check the flat space limit, we will set $f(\vec{x})=0$ and use standard QFT integrals to evaluate this (see, e.g., Appendix B.3.2 of \cite{Schwartz:2013pla}). Defining
\begin{align}\label{eq:Scaling}
    a&=\frac{\Delta u-i\epsilon}{(u-u_0-i\epsilon)(U-u_0+i\epsilon)}\,,\quad \vec{b}=\frac{(U-u_0+i\epsilon)\vec{x}-(u-u_0-i\epsilon)\vec{X}}{(u-u_0-i\epsilon)(U-u_0+i\epsilon)}\,,\nonumber\\
    c&=\Delta v-i \epsilon+\frac{(u-u_0-i\epsilon)\vec{X}^2-(U-u_0+i\epsilon)\vec{x}\,^2}{(u-u_0-i\epsilon)(U-u_0+i\epsilon)}\,,
\end{align}
we have
\begin{align}\label{eq:FSFinalInt}
    \int\frac{d\vec{x}\,'}{(2\pi)^{D-2}}\,&\left(\Delta v-i\epsilon- \frac{(\vec{x}-\vec{x}\,')^2}{u-u_0-i\epsilon}+\frac{(\vec{X}-\vec{x}\,')^2}{U-u_0+i\epsilon}\right)^{2-D}\notag\\
    %&=\int\frac{d\vec{x}\,'}{(2\pi)^{D-2}}\,\left[a(\vec{x}\,')^2+\frac{a c-\vec{b}\,^2}{a}\right]^{2-D}
    &=\int\frac{d\vec{x}\,'}{(2\pi)^{D-2}}\,\left(a \vec{x}\,'\,^2+2\vec{b}\cdot\vec{x}\,'+c\right)^{2-D}=\frac{\pi^\frac{D-2}{2}\Gamma(D/2-1)}{(2\pi)^{D-2}\Gamma(D-2)}(ac-\vec{b}^2)^{(2-D)/2}\nonumber\\
    &=\frac{\pi^\frac{D-2}{2}\Gamma(D/2-1)}{(2\pi)^{D-2}\Gamma(D-2)}\left(\frac{(\Delta v-i\epsilon)(\Delta u-i\epsilon)-(\vec{x}-\vec{X})^2}{(u-u_0-i\epsilon)(U-u_0+i\epsilon)}\right)^{(2-D)/2}\,.
\end{align}
The factor $(u-u_0)(U-u_0)$ cancels with the same prefactor in \eqref{eq:finalW_app}. Thus we get the flat space limit
\begin{align}
    W(u,v,\vec{x};U,V,\vec{X})=\frac{\Gamma(D/2-1)}{4\pi^{D/2}}\bigr[i\sigma(\mathsf{x},\mathsf{y})\bigr]^{(2-D)}\,,
\end{align}
where $\sigma({\sf x},{\sf y}) = \sqrt{-({\sf x-y})^2}=\sqrt{([\Delta u-i\epsilon][\Delta v-i\epsilon])-(\vec{x}-\vec{X})^2}$, recalling our convention ${\sf x}=(u,v,\vec{x})$ and ${\sf y}=(U,V,\vec{X})$. Indeed in $D=4$ this simply becomes
\begin{align*}
    W(u,v,\vec{x};U,V,\vec{X})=-\frac{1}{4\pi^2}\frac{1}{\sigma(x,y)^2} = -\frac{1}{4\pi^2}\frac{1}{(\Delta u-i\epsilon)(\Delta v-i\epsilon)-(\vec{x}-\vec{X})^2}\,.
\end{align*}

\subsection{Planar shockwave profile}\label{App:Gauss_SW}

Let us now calculate explicitly the Wightman function for the planar shockwave \eqref{eq:Gauss_profile}. Starting from \eqref{eq:finalW_app} and substituting the profile $f(\vec{x})=-\sum_i a_i (x^i)^2 $ we are left with the following integral:
\begin{align}
    &\int\frac{d\vec{x}\,'}{(2\pi)^{D-2}}\,\left(\Delta v-i\epsilon+\left(\sum_{i=1}^{D-2}a_i\Delta\Theta_u([x']^i)^2\right)- \frac{(\vec{x}-\vec{x}\,')^2}{u-u_0-i\epsilon}+\frac{(\vec{X}-\vec{x}\,')^2}{U-u_0+i\epsilon}\right)^{2-D}\notag\\
    &=\prod_{i=1}^{D-2}\left(1+a_i\Delta\Theta_u\frac{(u-u_0-i\epsilon)(U-u_0+i\epsilon)}{\Delta u-i\epsilon}\right)^{-1/2}\ \int\frac{d\vec{r}}{(2\pi)^{D-2}}\,\left(a \vec{r}\,^2+2\vec{\beta}\cdot\vec{r}+c\right)^{2-D}\,.
\end{align}
Here we have defined $a$, $\vec{b}$ and $c$ as in \eqref{eq:Scaling}, and
\begin{align}
    r^i&=\left(1+a_i\Delta\Theta_u\frac{(u-u_0-i\epsilon)(U-u_0+i\epsilon)}{\Delta u-i\epsilon}\right)^{1/2}(x')^i\,,\nonumber\\ 
     \beta^i&=\left(1+a_i\Delta\Theta_u\frac{(u-u_0-i\epsilon)(U-u_0+i\epsilon)}{\Delta u-i\epsilon}\right)^{-1/2} b^i\,,
\end{align}
where there is no summation over the indices. Thus we can directly apply \eqref{eq:FSFinalInt} up to the final simplification. Then we are left with the term $(ac-\vec{\beta}^2)$ which simplifies as follows:
\begin{align}
    &(u-u_0-i\epsilon)(U-u_0+i\epsilon)(ac-\vec{\beta}^2)\nonumber\\
    &=(\Delta v-i\epsilon)(\Delta u-i\epsilon)- \sum_{i=1}^{D-2}\left(1+a_i\Delta\Theta_u\frac{(u-u_0-i\epsilon)(U-u_0+i\epsilon)}{\Delta u-i\epsilon}\right)^{-1}\nonumber\\
     &\times\left[(x^i-y^i)^2-a_i \Delta\Theta_u\frac{(u-u_0-i\epsilon)(U-u_0+i\epsilon)}{\Delta u-i\epsilon} \left(\frac{(x^i)^2}{u-u_0-i\epsilon}-\frac{(y^i)^2}{U-u_0+i\epsilon}\right) \right]\nonumber\\
    &=(\Delta v-i\epsilon)(\Delta u-i\epsilon)-\Delta \vec{x}\,^2 + \sum_{i=1}^{D-2}\frac{a_i\Delta\Theta_u\left([U-u_0+i\epsilon]x^i-[u-u_0-i\epsilon]X^i\right)^2}{\Delta u+a_i\Delta\Theta_u(u-u_0-i\epsilon)(U-u_0+i\epsilon) }\,.
\end{align}
Finally, putting everything together, we obtain \eqref{eq: Wightman-Gauss} of the main text
\begin{multline}
     W(u,v,\vec{x};U,V,\vec{X})
    =\frac{(-i)^{D-2}\Gamma(D/2-1)}{4\pi^{D/2}} \prod_{i=1}^{D-2}\left(1+a_i\Delta\Theta_u\frac{(u-u_0-i\epsilon)(U-u_0+i\epsilon)}{\Delta u-i\epsilon}\right)^{-1/2}\nonumber\\
    \times\left((\Delta v-i\epsilon)(\Delta u-i\epsilon) -\Delta \vec{x}\,^2 \!+\! \sum_{i=1}^{D-2}\frac{a_i\Delta\Theta_u\left([U-u_0+i\epsilon]x^i-[u-u_0-i\epsilon]X^i\right)^2}{\Delta u+a_i\Delta\Theta_u(u-u_0-i\epsilon)(U-u_0+i\epsilon) }\right)^{\frac{2-D}{2}}\,.
\end{multline}

\section{Geodesics}
\label{app:B}

In this appendix we present a solution to the geodesic equations on the shockwave geometry. While this problem is now well understood from both physical and mathematical perspectives, see e.g. \cite{Dray1985shockwave,ferrari1988beam,balasin1997geodesics,kunzinger1999rigorous}, here we present a `simplified derivation' based on standard calculus, augmented with well-known properties of distributional derivatives and `natural' choice of products of distributions. For the choice of metric we are considering, it turns out that these `heuristic' calculations can be given rigorous justification, e.g.  \cite{kunzinger1999rigorous}.

Starting from the shockwave geometry in Brinkmann coordinates, \eqref{eq:metric}, the nonzero Christoffel symbols are
\begin{align}
    \Gamma^{v}_{uu} = -f\delta'(u-u_0)\,,\quad 
    \Gamma^{v}_{ui} = -\delta(u-u_0)\partial_i f\,,\quad     \Gamma^{i}_{uu} = -\frac{1}{2}\delta(u-u_0)\partial^if\,,
\end{align}
where $\delta'(u-u_0) $ is the distributional derivative of the delta function with respect to $u$. Denoting an affine parameter by $\tau$,
the geodesic equation yields
\begin{align}
    \frac{d ^2 u}{d \tau^2} &= 0 \,,\\
    \frac{d ^2 v}{d \tau^2} &= f\delta'(u-u_0)\rr{\frac{d  u}{d  \tau}}^2 +2\delta(u-u_0)\partial_if\frac{d  u}{d \tau}\frac{d  x^i}{d \tau}\,,\\
    \frac{d ^2 x^i}{d \tau^2} &= \frac{1}{2}\delta(u-u_0)\partial^if\rr{\frac{d  u}{d \tau}}^2\,,
\end{align}
subject to the norm constraint
\begin{align}
    g_{\mu\nu}\frac{d  x^\mu}{d \tau}\frac{d  x^\nu}{d \tau} = f\delta(u-u_0)\rr{\frac{d  u}{d \tau}}^2 + \frac{d  u}{d \tau}\frac{d  v}{d \tau} + \rr{\frac{d  x^i}{d \tau}}^2
    = e\,,
\end{align}
where $e=0,-1,1$ for null, timelike, spacelike geodesics, respectively. It follows that $u(\tau) = A\tau+B$ for some constants $A,B$, thus $u$ itself is a valid affine parameter for the geodesics. This allows us to rewrite $\frac{d}{d\tau}=A\frac{d}{du}$. 
To solve the remaining geodesic equations, we shall use the following properties of distributional derivatives:\footnote{Strictly speaking this is only a heuristic argument because we are not directly dealing with subtle issues with products of distributions very carefully. Instead, we will be integrating using rules of calculus but augmented with `distributional (anti-)derivatives' and reasonable assumptions on `mild' product of distributions, such as $\Theta(u-u_0)^2=\Theta(u-u_0)$.}
\begin{align}
    \frac{d }{d  u}\Theta(u-u_0) &= \delta(u-u_0)\,,\\
    \frac{d }{d  u}\left[(u-u_0)\Theta(u-u_0)\right] &= \Theta(u-u_0)\,.
\end{align}
By integrating the geodesic equation twice (taking into account `distributional' antiderivatives), we get
\begin{align}
    \frac{d  x^i}{d  u} &= \frac{1}{2}\Theta(u-u_0)\partial^if + C^i\,,\label{eq: velocity-transverse}\\
    x^i(u) &= C^i(u-u_0)+D^i + K^i(u-u_0)\Theta(u-u_0)\,,
    \label{eq: geodesic-xy}
\end{align}
where $C^i,D^i, K^i$ are integration constants that depend on initial data $x_0^i$ and the shockwave profile $f$; in particular $D^i=x_0$, $C^i=\dot{x}^i_0$, and $K^i=\frac{1}{2}\partial^if(x_0)$.

The solution for $v(u)$ is more complicated if we attempt to directly integrate the geodesic equation due to products of distributions (which are ill-defined outside of the Colombeau algebra of generalized distributions). To simplify things we use the norm constraint, since the equation of motion for $x^i(u)$ is distributionally well-defined. That is, we substitute \eqref{eq: velocity-transverse} into the norm constraint, to get
\begin{align}
    \frac{d  v}{d  u} &= \frac{e}{A^2}-f\delta(u-u_0)-\rr{\frac{1}{2}\Theta(u-u_0)\partial_if+C_i}^2\notag\\
    &= \frac{e}{A^2}-f\delta(u-u_0)- \frac{1}{4}(\partial_if)(\partial^if)\Theta(u-u_0)^2 - C^i\partial_if\Theta(u-u_0)-C_iC^i\,.
\end{align}
At this point we are again faced with products of distributions, however this is not difficult to deal with. A reasonable assumption would be to take $\Theta^2 = \Theta$ (which in calculus is a natural choice to make). By direct integration and collecting the constants together, we can express the solution as
\begin{align}
    v(u) = v_0 + \dot{v}_0(u-u_0) - f \Theta(u-u_0) - \left[C^i\partial_i f+ \frac{1}{4}(\partial_if)^2\right](u-u_0)\Theta(u-u_0)\,.
    \label{eq: geodesic-v}
\end{align}
Here $v_0$ is arbitrary and $\dot{v}_0=e/A^2-C_iC^i$. 

The main problem we need to address in this paper is the fact that, in the presence of the shockwave, one cannot in general pick a nice codimension-1 spacelike surface that serves as a simultaneity hyperplane because the spacetime does not admit a smooth spacelike Cauchy surface. The next best thing we can ask for is to see if by restricting observers (timelike geodesics) to be on a certain codimension-2 plane, we can at least have a `simultaneity curve' on this plane. On any other plane, the geodesics will be `refracted' by the shock (including the spacelike ones), complicating any attempts to measure proper distances between two events across the shock. 

\begin{figure}[h!]
    \centering
    \includegraphics[scale=0.775]{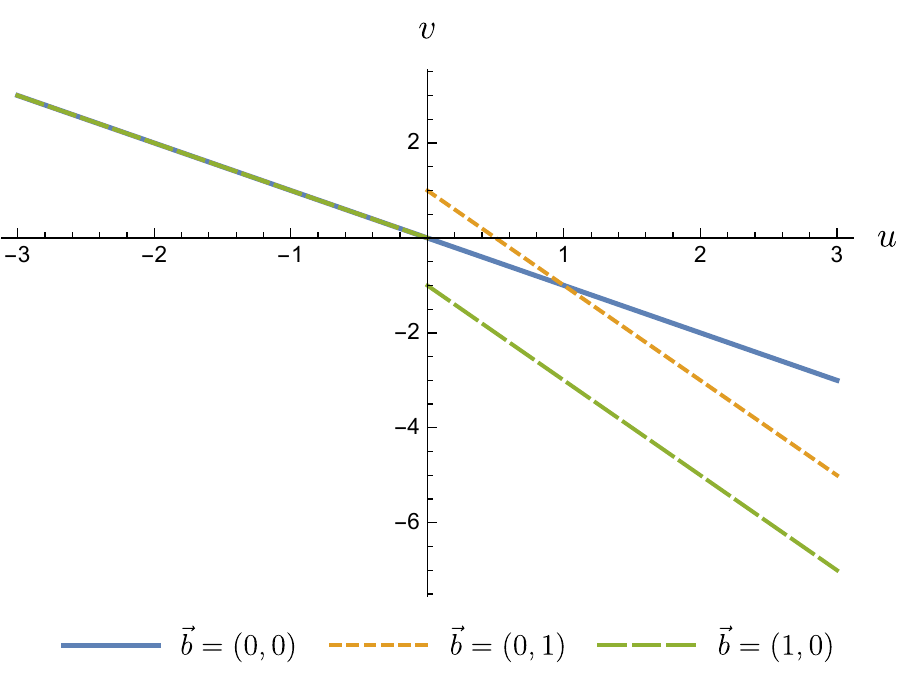}
    \includegraphics[scale=0.775]{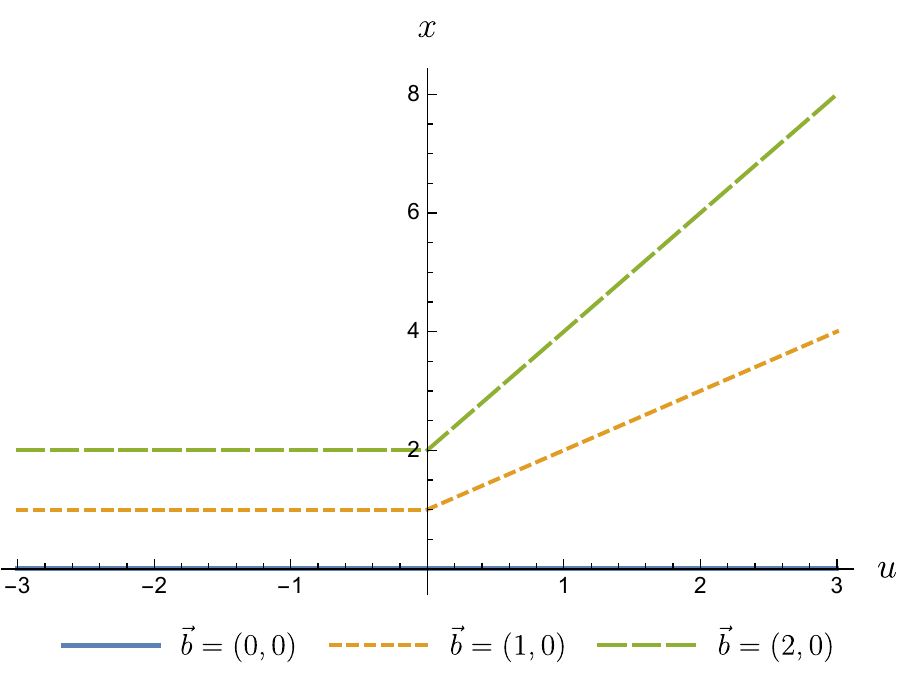}
    \caption{\textbf{Geodesics in shockwave spacetime.} We display timelike geodesics corresponding to a static trajectory at fixed $z_0=2T$ for different impact parameters $\vec{b}$. We consider the case where the trajectory has no initial velocity before encountering the shockwave; this corresponds to $dx/du = dy/du = 0$, so that $C^i=0$, and $dv/du = -1$ for $u<u_0$. For this trajectory the proper time $\tau$ is related to the affine parameter $u$ by a constant shift $u = \tau-z_0$. The plot for $y(u)$ is similar to the $x(u)$ plot on the right. Due to the choice of the shockwave profile and coordinate system $(u,v, \vec{x})$, we see that from the metric~\eqref{eq:metric} the static trajectories that lie in the $\vec b = 0$ plane are not affected by the shockwave.}
    \label{fig: geodesics-plane}
\end{figure}

It turns out that for our choice of planar shockwave given by the profile $f(\vec{x})=\sum_{i=1}^{D-2}a_i (x^i)^2$, we are very lucky. For this to work, we can choose the impact parameter of the trajectories of both detectors to be $\vec{b}=0$, corresponding to $x^i = 0$ for all $i=1,2,..., D-2$, then the contributions to the geodesics due to the shockwave for these trajectories \textit{vanish}:
\begin{align}
    f(\vec{b}=\vec{0}) = \partial_if\Bigr|_{\vec{b}=0} = 0\,.
\end{align}
What this means is that if we consider only timelike geodesics of the form
\begin{align}
    \sx(\tau) &= (u(\tau),v(\tau),\vec{0})\,,
\end{align}
which corresponds to trajectories constrained to the $(t,z)$-plane,
then the solution of the geodesic equation is precisely equal to the Minkowski one, independently of the shockwave profile. Therefore, the shockwave does not alter the behaviour of the geodesics on the plane $\vec{x} = 0$. We show how timelike geodesics can be altered by the impact parameter in Fig.~\ref{fig: geodesics-plane}; the results for other geodesics are similar by virtue of the solution \eqref{eq: geodesic-v} and \eqref{eq: geodesic-xy}.

In particular, it means that on this plane alone, we can take any two points labeled by $(t,z_A)$ and $(t,z_B)$, situated on different sides of the shockwave, and still argue that the proper separation is simply given by $|z_B-z_A|$ since the spacelike geodesic $t=\text{constant}$ is well-defined, thus defining a simultaneity curve on the subspace $(t,z,\vec{0})$.

\bibliography{shockrefs}
\bibliographystyle{JHEP}

\end{document}